\documentclass[lettersize,journal]{IEEEtran}
\usepackage{amsmath,amsfonts}
\usepackage{algorithmic}
\usepackage{algorithm}
\usepackage{array}
\usepackage[caption=false,font=normalsize,labelfont=sf,textfont=sf]{subfig}
\usepackage{textcomp}
\usepackage{stfloats}
\usepackage[table,xcdraw]{xcolor}

\usepackage{url}
\usepackage{url}

\usepackage{verbatim}
\usepackage{graphicx}
\usepackage{cite}

\usepackage{hyperref}
\usepackage{microtype}
\usepackage{graphicx}
\usepackage{subfig}
\usepackage{booktabs} % for professional tables
\usepackage{multirow}
\usepackage{multicol}
\usepackage{makecell}
\usepackage{amssymb}
\usepackage{mathtools}
\usepackage{amsthm}
\usepackage{bm}
\usepackage{xcolor}
\usepackage{subcaption}
 \usepackage{fancyhdr}

\fancypagestyle{mainFancy}{
    \fancyhf{}
    \fancyhead[C]{\fzkai{\leftmark}}       % 页眉章标题
    \fancyhead[C]{\sffamily \fontsize{8}{10}\selectfont This article has been accepted for publication in IEEE Journal of Selected Topics in Signal Processing. This is the author's version which has not been fully edited and
content may change prior to final publication. Citation information: 10.1109/JSTSP.2024.3488557}
    \fancyfoot[C]{\sffamily \fontsize{8}{10}\selectfont© 2024 IEEE. Personal use is permitted, but republication/redistribution requires IEEE permission.} % 将拇指放到没有被使用的页眉或页脚处
}
\begin{document}

%\title{UniCodec: Universal Neural Speech Codec Token based Framework for Speech Generation}
\title{Universal Speech Token Learning via Low-Bitrate Neural Codec and Pretrained Representations}
%\icmltitle{Universal Speech Token Learning via Low-Bitrate Neural Codec}
%\icmltitle{Universal Speech Token Learning for generative modeling}
\thispagestyle{mainFancy}

\author{Xue Jiang, Xiulian Peng, Yuan Zhang, Yan Lu%,~\IEEEmembership{~IEEE Senior Member}
        % <-this % stops a space
%\thanks{This paper was produced by the IEEE Publication Technology Group. They are in Piscataway, NJ.}% <-this % stops a space
%\thanks{Manuscript received April 19, 2021; revised August 16, 2021.}
\thanks{Xue Jiang is with the School of Information and Communication Engineering, Communication University of China, Beijing 100024, China (e-mail: jiangxhoho@cuc.edu.cn).}
\thanks{X. Peng and Y. Lu are with the Microsoft Research Asia, Beijing 100080, China (e-mail: xipe@microsoft.com; huxue@microsoft.com; yanlu@microsoft.com).}
\thanks{Y. Zhang is with the State Key Laboratory of Media Convergence and Communication, Communication University of China, Beijing 100024, China (e-mail: yzhang@cuc.edu.cn).}
\thanks{This work was done when Xue Jiang was an intern at Microsoft Research Asia.}}

%\author{IEEE Publication Technology,~\IEEEmembership{Staff,~IEEE,}
%        % <-this % stops a space
%\thanks{This paper was produced by the IEEE Publication Technology Group. They are in Piscataway, NJ.}% <-this % stops a space
%\thanks{Manuscript received April 19, 2021; revised August 16, 2021.}}

% \markboth{submitted to IEEE JSTSP Special Issue on Neural Speech and Audio Coding}%
% \markboth{\fontsize{8}{10}\selectfont This article has been accepted for publication in IEEE Journal of Selected Topics in Signal Processing. This is the author's version which has not been fully edited and
% content may change prior to final publication. Citation information: 10.1109/JSTSP.2024.3488557}

\IEEEpubid{0000--0000/00\$00.00~\copyright~2024 IEEE}

% \maketitle
\maketitle\thispagestyle{fancy}
\begin{abstract}
Current large speech language models are mainly based on semantic tokens from discretization of self-supervised learned representations and acoustic tokens from a neural codec, following a semantic-modeling and acoustic-synthesis paradigm. However, semantic tokens discard paralinguistic attributes of speakers that is important for natural spoken communication, while prompt-based acoustic synthesis from semantic tokens has limits in recovering paralinguistic details and suffers from robustness issues, especially when there are domain gaps between the prompt and the target. This paper unifies two types of tokens and proposes the UniCodec, a universal speech token learning that encapsulates all semantics of speech, including linguistic and paralinguistic information, into a compact and semantically-disentangled unified token. Such a unified token can not only benefit speech language models in understanding with paralinguistic hints but also help speech generation with high-quality output. A low-bitrate neural codec is leveraged to learn such disentangled discrete representations at global and local scales, with knowledge distilled from self-supervised learned features. Extensive evaluations on multilingual datasets demonstrate its effectiveness in generating natural, expressive and long-term consistent output quality with paralinguistic attributes well preserved in several speech processing tasks. 
% \footnote{ Audio samples are available at \href{https://unicodec2.github.io/demo/}{https://unicodec2.github.io/demo/.}}.
\end{abstract}

\begin{IEEEkeywords}
Speech language model, neural speech codec, speech synthesis.
\end{IEEEkeywords}

\section{Introduction}

\begin{figure*}[htb] 
\centering 
\includegraphics[width=1.0\textwidth]{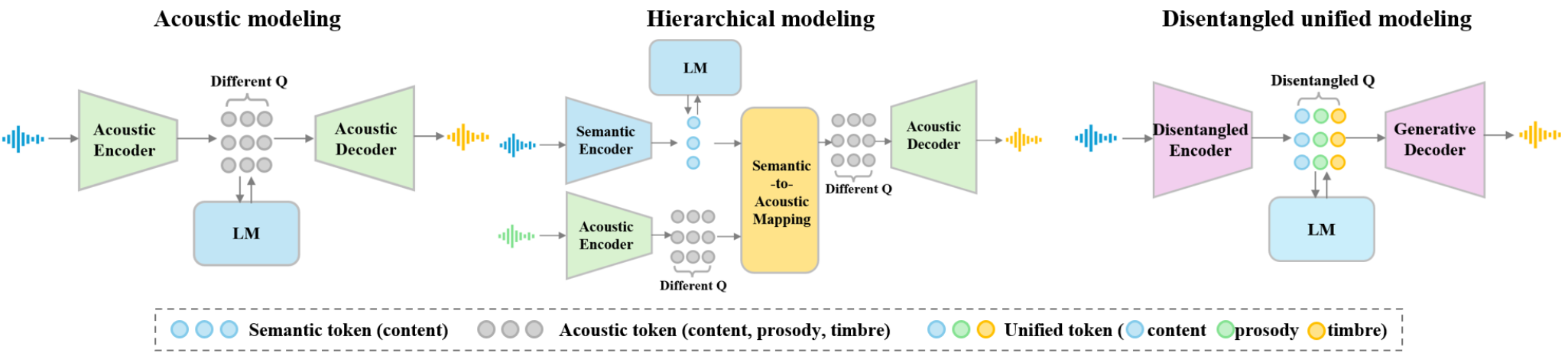}
\vspace{-0.2cm}
\caption{Illustration of different types of discrete speech representations and their typical usage in generative speech language modeling. Our unified token encapsulates disentangled semantics into a compact token. LM: language modeling.}
\vspace{-0.4cm}
\label{Fig_highlight} 
\end{figure*}

Recently we have witnessed substantial breakthrough in large language models (LLMs), demonstrating exceptional abilities in language understanding, reasoning, and more. Compared with text, speech is the more natural form of communication for human interactions. Beyond mere words, speech carries a wealth of additional information that text alone cannot capture, i.e. paralinguistic information of human attributes such as speaking styles, timbre, emotion, sentiment, etc. These information helps in how we perceive and understand each other, including not just what a person says but who says it and how. Different ways of speaking the same sentence may have completely different meanings. Therefore, it is essential to model both linguistic and paralinguistic information of speech.

The progress in large language models has led to the emergence of many speech unit based algorithms that are based on discrete speech tokens rather than continuous representations, where a speech utterance is formulated as a discrete token sequence, similar to that in text. This formulation makes it feasible to leverage knowledge and techniques in textual LLMs for speech processing and promising results have been achieved along this line in direct speech-to-speech translation (S2ST) \cite{AudioPaLM}, automatic speech recognition (ASR) \cite{speech-gpt, AudioPaLM}, spoken dialogue \cite{speech-gpt}, etc. They typically take a semantic speech modeling approach \cite{GSLM, speech-gpt, AudioPaLM}, leveraging a so-called semantic token from a self-supervised learned feature with k-means clustering such as wav2vec 2.0 \cite{wav2vec2} and HuBERT \cite{HuBERT}. However, the semantic token primarily encodes linguistic information, often overlooking the nuances of paralinguistic cues that is important in natural spoken communication \cite{Prosodic, Para-dialogue, pGSLM}.

Recently, the acoustic token from a neural codec (e.g. SoundStream \cite{SoundStream} and EnCodec \cite{EnCodec}) is proposed, which draws a lot of attention in speech language models (LM) \cite{AudioLM, valle, SPEAR-TTS, LM-VC}. In this line, there are two typical generative speech modeling paradigms, the pure acoustic token based \cite{valle, gen-dns} and the hierarchical language modeling (Hierarchical LM) \cite{AudioLM, SPEAR-TTS, LM-VC} that leverages both semantic and acoustic tokens (see Fig. \ref{Fig_highlight}). The acoustic token paradigm gives high reconstruction quality but it lacks language modeling capability without a semantic representation as indicated in AudioLM \cite{AudioLM}. Moreover, it typically gives longer token lengths than semantic ones, which affects the generation complexity and robustness. Phonetic errors are reported in text-to-speech synthesis \cite{valle}. Comparatively, the Hierarchical LM approach takes advantages of both types of tokens and achieves good results in both speech language modeling and high-quality speech synthesis \cite{AudioPaLM, SPEAR-TTS}. However, its language modeling is purely based on linguistic part, losing paralinguistic context in understanding. What's more, in speech synthesis, the semantic-to-acoustic mapping simply adds back local prosody and other paralinguistic information through in-context learning from a prompt with acoustic tokens. This prompt-based method has limits in synthesizing local details such as intonation that carries rich paralinguistic information in a fine manner, especially when there are domain gaps between the prompt and the target such as cross-lingual cases. 
%Also there are robustness issues when synthesizing those information from the prompt in an autoregressive way. 
\IEEEpubidadjcol

In light of this, this paper proposes a universal speech token learning paradigm based on low-bitrate neural codec and self-supervised learning (SSL) representations, named UniCodec, for prosody-aware speech language modeling. Taking reconstruction as a self-supervised objective, our low-bitrate UniCodec employs a \textbf{\textit{compression-generation}} framework (see Fig.\ref{Fig_highlight}) to compress key attributes of speech, including both linguistic and paralinguistic information, into compact discrete tokens that are good for prosody-aware speech modeling, and still deliver high reconstruction quality by a generative decoder. Compared with acoustic tokens, the UniCodec tokens are much more compact and easy to predict in speech language models, while compared with Hierarchical LM, it can achieve more natural, robust, and expressive speech generation by incorporating paralinguistic information into speech language models. In this sense, we turn the low bitrate neural codec model into a powerful discrete representation learner that captures both linguistic and paralinguistic information in a compact and unified representation, with information distilled from SSL representations.

The main contributions of this paper are summarized as follows.
\begin{itemize}
\item We unify semantic and acoustic tokens into a compact universal token with short token lengths for prosody-aware generative speech language modeling. We introduce a low-bitrate neural codec framework to learn this disentangled discrete tokens at both global and local levels, with knowledge from SSL representations.
\item We design extensive experiments on multilingual datasets to validate the representation ability of UniCodec tokens in speech understanding and emotion recognition, and its ability of paralinguistic-aware speech language modeling in text-to-speech synthesis and speech-to-speech translation tasks. 
\item We make a thorough comparison with Hierarchical LM and acoustic token based approaches and show that the UniCodec token achieves good performance across a range of tasks by efficiently integrating linguistic and paralinguistic cues in a compact and predictable manner. 
\end{itemize}

\begin{figure*}[htb] 
\centering 
\includegraphics[width=1.0\textwidth]{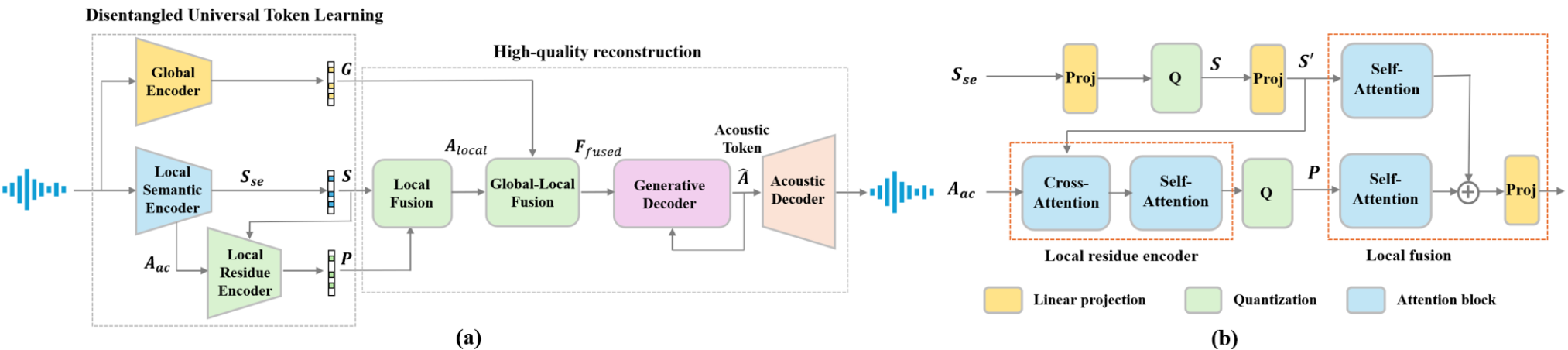} %MF-Audio2
\caption{(a) The proposed framework. (b) Local disentangled encoders and local fusion.}
\label{Fig_proposed_framework} 
\end{figure*}

%\begin{figure*}[htb] 
%\centering 
%\includegraphics[width=0.85\textwidth]{figure/MF-Audio2.png}
%\caption{The proposed framework.}
%\label{Fig_proposed_framework} 
%\end{figure*}

%\begin{figure}[tb] 
%\centering 
%\includegraphics[width=0.87\columnwidth]{figure/extractor2.png}
%\vspace{-0.3cm}
%\caption{Local disentangled encoders and local fusion.}
%\vspace{-0.3cm}
%\label{Fig_local} 
%\end{figure}

\section{Related Work}
%\ {
\subsection{Speech Representation Learning}
Self-supervised learning (SSL) methods have demonstrated impressive performance in recent years \cite{wav2vec2, HuBERT}. They mainly focus on extracting semantic information from speech through various pretraining tasks such as masked prediction and contrastive learning, achieving good performance in automatic speech recognition. However, the paralinguistic information which plays a crucial role in both understanding and high quality speech generation are often discarded, especially after k-means clustering to get the semantic tokens introduced above. Therefore, to learn more universal discrete representations, we introduce the low bitrate neural codec with a reconstruction task to help retain paralinguistic information in the learned tokens. Notably, the generative decoder utilized in our reconstruction task is jointly trained with the discrete representations, making the learned tokens more suitable for high-quality speech generation.
%}
\subsection{Generative Speech Language Modeling}
Discrete speech representations have become a major basis to build generative speech language modeling recently. Lakhotia et al. proposed the GSLM \cite{GSLM} for 'textless NLP' that leverages a semantic token from a self-supervised representation learning feature with k-means clustering such as wav2vec 2.0 \cite{wav2vec2} and HuBERT \cite{HuBERT}. Such a paradigm is also employed to build large foundation models for speech as in SpeechGPT \cite{speech-gpt}. As its reconstruction loses paralinguistic information, Borsos et al. further proposed the AudioLM \cite{AudioLM}, a hierarchical language modeling approach based on both semantic tokens and acoustic tokens from the SoundStream neural codec \cite{SoundStream}, which achieves high audio quality and long-term consistency in speech continuation with three-stage token generation. Similar hierarchical modeling is further explored in text-to-speech synthesis (TTS) \cite{SPEAR-TTS}, voice conversion \cite{LM-VC} and large foundation models \cite{AudioPaLM}. Thanks to the high-quality synthesis of speech, there are some works that use acoustic tokens only. With text to provide linguistic condition, the VALL-E \cite{valle} introduces a conditional language modeling to generate acoustic tokens of speech for TTS. This is demonstrated to be inferior to hierarchical modeling approach in SPEAR-TTS \cite{SPEAR-TTS}. The acoustic modeling is also investigated in other tasks such as speech enhancement \cite{gen-dns}. Compared with these tokens and speech modeling paradigms, our approach provides a unified token representation that disentangles various semantics of speech in a compact representation.
 
%pGSLM, X-LLM,, SpeechLM, SPECTRON, SeamlessM4T, LTU
\vspace{-0.1cm}
\subsection{Neural Speech Codecs}
\vspace{-0.1cm}
Currently popular neural codecs mainly adopt the VQVAE pipeline \cite{VQ-VAE,EnCodec,SoundStream}, where an encoder, a vector quantizer, and a decoder are jointly optimized in an end-to-end fashion. The discrete representation by vector quantization provides acoustic tokens that carries all acoustic details to reconstruct the waveform in a high fidelity. There are two typical types of acoustic tokens, one by residual vector quantization (RVQ) and the other by group vector quantization (GVQ). In RVQ, the features are quantized in a coarse-to-fine fashion and multiple stages are needed in generating acoustic tokens when used in speech synthesis \cite{AudioLM, valle}. In GVQ, different groups are quantized independently and thus a single stage is feasible to generate all groups of acoustic codes simultaneously for each frame \cite{gen-dns}. The SpeechTokenizer \cite{SpeechTokenizer} shares somewhat similar motivation with our work, but its tokens are still based on a coarse-to-fine style including semantic content tokens followed by acoustic details; so it somewhat resembles the hierarchical language modeling. Compared with acoustic tokens, our universal tokens are much more compact and semantic-rich with disentanglement; thus it is more easy to predict in speech generation. On the other hand, as it encapsulates various key attributes of speech, such as linguistic and paralinguistic information captured at global and local scales, a high quality synthesis with high perceptual fidelity can still be achieved.

\vspace{-0.1cm}
\subsection{Paralinguistics-Enhanced Speech Processing}
Paralinguistic information plays an important role in achieving natural communication. Lin et al. introduced a global sentiment embedding for spoken dialogues which achieves better response text prediction \cite{Para-dialogue}. Ward et al. introduced prosodic clues for spoken diaglogue \cite{Prosodic}. Kharitonov et al. proposed to add extra pitch and duration units to improve speech naturalness \cite{pGSLM}. All these methods have demonstrated the importance of paralinguistic information; however, they either use handcrafted or explicitly modeled features. Our UniCodec leverages a neural codec with high-fidelity reconstruction to ensure that all semantics or attributes of speech are captured in the disentangled tokens and all details are well reconstructed given the discrete token. Moreover, we model this information at both global and local scales as some information such as emotion has both global representation and local component dependent on content.  

\vspace{-0.2cm}
\section{The Proposed UniCodec}
%\label{submission}

% Prior approaches to learning representations of prosody have relied on subtractive definitions such as "Prosody is the variation
% in speech signals that remains after accounting for variation due to phonetics, speaker identity, and channel effects"
\subsection{Problem Formulation}
Let $x$ denote the speech waveform. The goal of our UniCodec is to encode $x$ into several information-disentangled discrete tokens, and resynthesis $x$ from these tokens, given a low bitrate constraint $R_{target}$ on the encoded tokens. It is denotes as
\begin{equation}
\begin{aligned}
{\bm{G},\bm{S},\bm{P}} &= {Q(UniCodecEnc}(x)),\\
\hat{x} &= {UniCodecDec} {(\bm{G},\bm{S},\bm{P})},
\end{aligned}
\end{equation}
where $UniCodecEnc$, $Q$ and $UniCodecDec$ are encoder, quantizer and decoder, respectively. According to speech characteristics that some information is more global while some is more rapidly changing, we disentangle $x$ at global and local scales into three levels of tokens ${(\bm{G},\bm{S},\bm{P})}$. $\bm{G}$ denotes a global token capturing \ {attributes that tend to remain consistent or stable across long stretches of speech}, such as timbre, speaking styles, emotion, volume, etc. $\bm{S}$ denotes local semantic tokens capturing linguistic content of speech. $\bm{P}$ denotes local residue tokens that carries residue information not included in global and local semantic tokens, e.g. prosody. This part $\bm{P}$ is important in carrying paralinguistic information such as emotion and sentiment for understanding and speech resynthesis without information loss. 

The UniCodec employs a \textbf{\textit{compression-generation}} framework with ${(\bm{G},\bm{S},\bm{P})}$ encoded at an extremely low bitrate $R_{target}$ and a strong generative decoder is leveraged for $UniCodecDec$ to deliver a high-quality reconstruction quality. It is optimized with a reconstruction objective under a low-bitrate constraint, denoted as
\begin{equation} %##(?)
\begin{aligned}
\mathcal{L} = \min\limits_{\theta} {\rm Distortion}(x,\hat{x})+\lambda_r |R_{target}-R(\bm{G},\bm{S},\bm{P})|,
\end{aligned}
\end{equation}
where $R(\bm{G},\bm{S},\bm{P})$ denotes the bits consumed by the representations. The basic idea behind this approach is that if the model can effectively resynthesis the original signal through those tokens, the learned tokens should encompass all crucial information present in the original speech, including high-level semantic information and fine prosody, without losing any linguistic and paralinguistic information in speech. Moreover, the low bitrate serves as an information bottleneck to achieve compactness of the learned semantic-rich tokens, making it more easy to predict in speech language models. The UniCodec token is optimized with the strong generative decoder jointly in an end-to-end fashion, leading to a more compact discrete representation learning. 

For better representation of different aspects of speech, disentanglement is encouraged for UniCodec learning through several techniques. First, the encoders for $\bm{G}$ and ${(\bm{S},\bm{P})}$ are global and local, encouraging them to capture information at different scales. Second, we incorporate pretrained representations to help learn semantic information in $\bm{S}$ with a semantic supervision explained later. Finally, the low bitrate information bottleneck encourages the disentanglement between $\bm{S}$ and $\bm{P}$. The compactness, disentanglement and the semantic-rich properties make the triplet token ${(\bm{G},\bm{S},\bm{P})}$ a good representation of speech for various speech understanding, conversion and generation tasks.

\vspace{-0.1cm}
\subsection{UniCodec}
\vspace{-0.1cm}
As illustrated in Fig. \ref{Fig_proposed_framework} (a), our UniCodec
follows an encoder-VQ-decoder framework, which takes the time-domain audio signal $x$ as input, produce the global tokens $\bm{G}$, local semantic tokens $\bm{S}$ and residual tokens $\bm{P}$ after vector quantization (VQ). The  decoder fuses these information-disentangled features and synthesis the speech $\hat{x}$. 

%\begin{itemize}
%\item 
\subsubsection{Encoder} As explained above, UniCodec introduces three encoders to decompose the speech components, which are the global encoder  $E_{g}$, local semantic encoder $E_{s}$ and local residual encoder $E_{p}$, formulated as
\begin{equation}
\begin{aligned}
{\bm{G}}=Q_g(E_{g}(x)),
{\bm{S}}=Q_s(E_{s}(x)),
{\bm{P}}=Q_p(E_{p}(x,S)).
\end{aligned}
\end{equation}
$E_{g}$ generates nearly time-invariant tokens $\bm{G}$ that encapsulate global information, \ {as stated in the previous section}. $E_{s}$ and $E_{p}$ produce frame-level tokens capturing content information and content-dependent local residues carrying rich paralinguistic information, respectively. 

For $E_{g}$, we use a speaker encoder similar to that in Disen-TF-Codec \cite{Disen-TF-Codec}, a self-supervised low-latency neural codec that disentangles global with local features. Unlike \cite{Disen-TF-Codec}, we further introduce some attention modules to better capture the global attributes adaptively. Specifically, our encoder takes time-frequency spectrum by short-time Fourier transform (STFT) as input and leverages \ {five} 2D convolutional blocks, \ {a reshape to fuse frequency into channel dimension, four} temporal filtering blocks and a \ {2-layer} transformer encoder to capture both short-term and long-term correlations. Finally, a multi-head attention with a learnable query followed by a linear head is leveraged to adaptively aggregate all frames into a global embedding \ {$\bm{G}\in \mathbb{R}^{1 \times C_{g}}$, where $C_{g}$ denotes the channel dimension. The details can be found in the appendix.} 

Instead of training from scratch, we leverage a self-supervised learned semantic encoder for $E_{s}$ and extract residue features through another $E_{p}$ conditioned on intermediate features of $E_{s}$ and $\bm{S}$ (see Fig. \ref{Fig_proposed_framework}). Specifically, we initialize $E_{s}$ from a pretrained multilingual HuBERT \cite{HuBERT}, i.e. mHuBERT Base \footnote{\href{https://dl.fbaipublicfiles.com/hubert/mhubert_base_vp_en_es_fr_it3.pt}{https://dl.fbaipublicfiles.com/hubert/mhubert\_base\_vp\_en\_es\_fr\_it3.pt}} trained on a 100K subset of the large multilingual corpus VoxPopuli. As the bottom or middle layers of HuBERT capture more acoustic information of speech and the top layers tend to capture more semantic information, we take the $11$th layer output of mHuBERT as our semantic feature $\bm{S}_{se} \ {\in \mathbb{R}^{T\times C_{a}}}$ as that in \cite{TextlessS2ST_realdata} and the $6$th layer of mHuBERT as our acoustic feature $\bm{A}_{ac} \ {\in \mathbb{R}^{T\times C_{a}}}$ to learn the local pair \ {$(\bm{S},\bm{P})$, $\bm{S}\in \mathbb{R}^{T\times C_{s}},\bm{P}\in \mathbb{R}^{T\times C_{p}}$.} \ {Here $C_{a}$, $C_{s}$ and $C_{p}$ are channel dimensions and $T$ is the number of $40$\textit{ms} frames. As our quantization is performed on a $40$\textit{ms} frame to produce a more compact token sequence while mHuBERT features are in a $20$\textit{ms} basis, a reshape to $40$\textit{ms} by stacking two adjacent frames is conducted before extracting $\bm{S}_{se}$ and $\bm{A}_{ac}$.} We should emphasize that as our focus is not on how to learn semantic features, but instead on investigating universal tokens capturing various attributes of speech with compactness and disentanglement that can benefit speech language modeling, here we just use existing mHuBERT in semantic encoder.

\ {As shown in Fig. \ref{Fig_proposed_framework} (b), the discrete semantic feature $\bm{S}$ is produced by a quantization after a linear projection of $\bm{S}_{se}$.} The local residue encoder takes the quantized semantic feature $\bm{S}$ and the $\bm{A}_{ac}$ as input to extract local residual information. As $\bm{S}$ and $\bm{A}_{ac}$ are in different feature space, we leverage a channel-wise cross-attention module followed by a channel-wise self-attention block to extract residues (see Fig. \ref{Fig_proposed_framework} (b)). In cross-attention, we extract query from \ {$\bm{S'} \in \mathbb{R}^{T\times C_{a}}$, the linear projected feature of $\bm{S}$}, key and value from $\bm{A}_{ac} \in \mathbb{R}^{T\times C_{a}}$ and the attended output \ {$\bm{A_{att} \in \mathbb{R}^{T\times C_{a}}}$} is given by  
 \begin{equation}
 \begin{aligned}
 \bm{Q} &=& Reshape(\bm{S'} W_{Q}) \in \mathbb{R}^{T\times\frac{C_a}{F}\times F},\\
 \bm{K} &=& Reshape(\bm{A}_{ac} W_{K}) \in \mathbb{R}^{T\times\frac{C_a}{F}\times F},\\
 \bm{V} &=& Reshape(\bm{A}_{ac} W_{V}) \in \mathbb{R}^{T\times\frac{C_a}{F}\times F},\\
 \bm{A_{att}}&=&Reshape(Softmax(\bm{Q}\cdot \bm{K}^{\mathsf{T}}/\sqrt{F})\bm{V}),
 \end{aligned}
\label{Eq_crossatt}
 \end{equation} 
 \begin{equation}
 \begin{aligned}
     \bm{A_{res}}&=\bm{A}_{ac}\hspace{-1mm}-\hspace{-1mm}\bm{A_{att}},
 \end{aligned}
 \label{Eq_crossatt_skip}
 \vspace{-0.5em}
 \end{equation}
 where $W_{Q} \in \mathbb{R}^{C_{a}\times C_a}$, $W_{K}\in \mathbb{R}^{C_a\times C_a}$, $W_{V}\in \mathbb{R}^{C_a\times C_a}$ are learnable matrices for $\bm{Q}$, $\bm{K}$ and $\bm{V}$, respectively. \ {For low-memory cost, we reshape each frame from $\mathbb{R}^{T\times{C_a}}$ to $\mathbb{R}^{T\times\frac{C_a}{F}\times F}$ with $\frac{C_a}{F}$ groups along the channel dimension and each group having a feature dimension of $F$. Then the attention is performed among groups to get the attended output \ {$\bm{A_{att}}$ with another reshape from $\mathbb{R}^{T\times\frac{C_a}{F}\times F}$ to $\mathbb{R}^{T\times{C_a}}$}. $F$ is set to 32 in our experiments.} An intermediate residual $\bm{A_{res}}\ {\in \mathbb{R}^{T\times C_a}}$ is obtained by a skip connection from $\bm{A}_{ac}$ to the attended output \ {$\bm{A_{att}}$} by subtraction rather than addition as shown in Eq. (\ref{Eq_crossatt_skip}). After that, a self-attention is further conducted on $\bm{A_{res}}$ to extract residue features $\bm{P_{res}}$ for quantization into $\bm{P}\ {\in \mathbb{R}^{T\times C_p}}$. This self-attention block is similar to Eq. (\ref{Eq_crossatt}), except that all $\bm{Q}$, $\bm{K}$ and $\bm{V}$ come from ${\bm{A}_{res}}$ and the output dimension is $C_p$ instead of $C_a$.   

%\item 
\subsubsection{Vector Quantization with Rate Control} We leverage the Distance-Gumbel-Softmax-based VQ method proposed in \cite{TF-Codec} for group vector quantizer training and rate control. Each frame of $\bm{S}$ and $\bm{P}$ is split into $M$ groups \ {along the channel dimension}, and each group is quantized with a separate codebook with $K$ codewords $\{e_1,e_2,...,e_K\} \in \mathbb{R}^{\frac{C_d}{M}\times K}$, where $C_d$ is the feature dimension of each frame to be quantized, i.e. $C_s$ for $\bm{S}$ and $C_p$ for $\bm{P}$. We set the target entropy for $\bm{S}$ and $\bm{P}$ to be $R^s_{target}$ and $R^p_{target}$, respectively, and those target bitrate is equally distributed among groups. As that in \cite{TF-Codec}, each codebook size is set to be larger than the target bitrate so that it allows unequal usage of different codewords within each codebook to capture the real distribution of the learned features. To achieve the target bitrate, we leverage the following rate constraint given by
\begin{equation}
\begin{aligned}
\mathcal{L}_{rate} &= |R^s_{target} - \mathcal{H}(\bm{S})|+|R^p_{target} - \mathcal{H}(\bm{P})|,
\end{aligned}
\end{equation}
where $\mathcal{H}(\cdot)$ is the estimated entropy as that in \cite{TF-Codec}.

%\item 
\subsubsection{Decoder} \ {The UniCodec decoder $UniCodecDec$ consists of the following three components:
\begin{itemize}
\item {The decoder fusion module $D_f$, which merges the three tokens $\bm{G}$, $\bm{S}$ and $\bm{P}$ at global and local scales and get the fused feature $\bm{F}_{fused}=D_f(\bm{G},\bm{S},\bm{P})$.}
\item {A generative decoder $D_g$, which autoregressively generates an acoustic token sequence $\bm{A}$ conditioned on $\bm{F}_{fused}$, optimized by maximizing the likelihood $p(\bm{A}|F_{fused};\theta_{D_{g}})$, where $\theta_{D_{g}}$ denotes model parameters of $D_{g}$. It produces an output sequence $\hat{\bm{A}}=D_g(\bm{F}_{fused})$.}
% A decoder-only Transformer language model that operates
% on the discrete acoustic tokens $A$, trained to maximize the likelihood
% QTt=1 p(ht|h<t). At inference time, the model predicts the token sequence $\hat{A}$ autoregressively}
\item {An acoustic decoder $D_a$, which converts the generated acoustic tokens $\hat{\bm{A}}$ back into the waveform, as $\hat{x} = D_{a}(\hat{\bm{A}})$}.
\end{itemize}
}

\paragraph{Global and Local Fusion}
\ {We first leverage a local fusion module to fuse the two local tokens $\bm{S}$ and $\bm{P}$}. Two separate self-attention blocks are employed to transform $\bm{S}$ and $\bm{P}$ into the same space and the transformed features are then added together, followed by a linear projection to get the fused local feature $\bm{A}_{local}$, \ {as shown in Fig. \ref{Fig_proposed_framework} (b). Similar channel-wise attention as that in Eq. (\ref{Eq_crossatt}) is employed here for self-attention.} Then the global token $\bm{G}$ is fused with $\bm{A}_{local}$ \ {(see Fig. \ref{Fig_proposed_framework} (a))} through an adaptive channel modulation, where a scale vector $\lambda$ and a bias vector $\beta$ are learned from $\bm{G}$ by two separate multi-layer perceptrons and the output feature $\bm{F}_{fused}$ is obtained by $\bm{F}_{fused}=\lambda \cdot \bm{A}_{local}+\beta$. 

\paragraph{Generative Decoder}
As a strong decoder would lead to more information compactness of the learned tokens, we leverage a generative decoder to help reconstruct the original signal from the fused feature. Motivated by the success of transformer in sequence-to-sequence tasks \ {\cite{attention}}, we formulate the generative decoder as a conditional next-token prediction problem in an autoregressive way based on a \ {decoder-only transformer}. It gets the fused feature $\bm{F}_{fused}$ as condition, and autoregressively generate the acoustic token $\bm{A}$ from a pretrained higher-bitrate neural codec TF-Codec\cite{TF-Codec} with high quality reconstruction.  \ {In the following, we will introduce the acoustic token sequence $\bm{A}$ extracted with TF-Codec, which serves as the target for the generative decoder module, followed by a detailed explanation of the autoregressive generation process.}
 %我们会先介绍用tf-codec提取的acoustic token，之后再介绍自回归generation的过程
 % The acoustic decoder $D_{a}$ then converts the generated acoustic tokens $\hat{\bm{A}}$ back into waveform, as $\hat{x} = D_{a}(\hat{\bm{A}})$. Details of each part will be explained in the following paragraphs.

 % It should be noted that although our generative decoder shares some similarity with the semantic-to-acoustic mapping that is used in Hierarchical LM, the condition in our algorithm contains rich information of speech, including not only linguistic but also local residues with paralinguistic cues that is not easy to be predicted by $\theta_{D_{g}}$ from $\bm{S}$ only; therefore our decoder is much more robust for speech synthesis even when no speech prompt is available.
%\end{itemize}

\textbf{Acoustic Tokens with TF-Codec} 
We use the pretrained TF-Codec \cite{TF-Codec} \footnote{  \url{https://github.com/microsoft/TF-Codec}}, a state-of-the-art low-latency neural speech codec, to extract the acoustic token sequence $\bm{A}$. TF-Codec follows a convolutional encoder-VQ-decoder paradigm, operating in the time-frequency domain. For simplicity, we disable the predictive loop in TF-Codec and use the non-predictive pretrained TF-Codec at 6 kbps that achieves high-fidelity speech reconstruction. Unlike RVQ, it 
% stacks the features to $50$ Hz and 
uses group vector quantization \ {(GVQ) where the input features stacked at 50 Hz are split into $G_a = 16$ groups along the channel dimension and} each group is quantized by a separate vector quantizer with $1024$ codewords,
% The quantized $G_a$ codes of each frame are concatenated,
 \ {yielding the acoustic token sequence $\bm{A} = [\bm{a}_0,\bm{a}_1,...,\bm{a}_{T_a-1}]^{\mathsf{T}} \in \mathbb{R}^{ T_a \times G_a}$. $\bm{a}_t=[a_{t}^0,a_{t}^1,...,a_{t}^{G_a-1}]^{\mathsf{T}}$ denotes the acoustic codes at $t$-th frame with $G_a$ groups and each element in $\bm{a}_t$ belongs to the set $\{0,1,2,...,1023\}$. $T_a$ is the number of acoustic frames.}

% \begin{equation}
% \begin{aligned}
% {\bm{\hat{A}}} &= GenerativeDec(\bm{G},\bm{S},\bm{P})\\
% {{\hat{x}}} &= {AcousticDec(\bm{\hat{A}}))
% \end{aligned}
% \end{equation}

\textbf{Single-stage Autoregressive Generation}  
% Different from RVQ that each quantization is dependent on previous stages of quantization and therefore multiple successive stages are needed in the previous work \cite{AudioLM,valle} to generate acoustic tokens of all stages in a coarse-to-fine manner. \ {Thanks to GVQ in TF-Codec that encodes each group independently, all $G_a$ codes in $\bm{a}_t$ can be generated simultaneously in a single stage. }
\ {As GVQ in TF-Codec encodes each group independently, all $G_a$ codes in $\bm{a}_t$ can be generated simultaneously in a single stage. This differs significantly from previous methods \cite{AudioLM,valle} that rely on residual vector quantization based acoustic tokens,} where each quantization stage depends on all previous ones, requiring multiple successive stages to generate acoustic tokens in a coarse-to-fine manner.
We leverage a causal autoregressive transformer decoder as the generative model for the single-stage acoustic token generation, which maximizes the \ {joint probability of the acoustic token sequence $\bm{A}$, i.e.}
\begin{equation}
\begin{aligned}
    p(\bm{A}|F_{fused}; \theta_{D_{g}})=\prod \limits_{t=0}^{T_a} p(\bm{a}_{t}|\bm{a}_{<t},F_{fused}; \theta_{D_{g}}),
\end{aligned}
\label{eq-decoder}
\vspace{-0.2cm}
\end{equation}
 where 
 %$\bm{a}_t=[a_{t}^0,a_{t}^1,...,a_{t}^{G_a-1}]$ is the $t$-th frame token of $\bm{A}$.
 $\bm{a}_{-1}$ is the $start$ token and $\bm{a}_{T_a}$ is the $eos$ token. \ {$\bm{a}_{<t}$ denotes all past frames from $\bm{a}_{-1}$ to $\bm{a}_{t-1}$ and we follow this notation in all equations below if not otherwise specified.}
To achieve this, the $G_a$ embeddings corresponding to $G_a$ codebooks are concatenated \ {along the channel dimension} as the transformer input and $G_a$ classification heads are employed to predict the $G_a$ acoustic codes \ {in $\bm{a}_t$} jointly by projecting the transformer's final hidden state to $G_a$ sets of logits, one for each of the $G_a$ codebooks. \ {It is noted that as the number of frames for $\bm{A}$ and $F_{fused}$ are different, i.e. $T_a=2 T$, a reshape of $F_{fused}$ from $\mathbb{R}^{T\times{C_a}}$ to $\mathbb{R}^{T_a\times\frac{C_a}{2}}$ is conducted before being taken as the condition in the transform decoder.}
\ {\paragraph{Acoustic Decoder} The decoder of the pretrained TF-Codec is used as the acoustic decoder to reconstruct the audio waveform from the generated acoustic tokens $\bm{\hat{A}}$. The acoustic decoder is freezed during our UniCodec training. \ {More details of the UniCodec configuration can be found in Appendix A.}}

\vspace{-0.2cm}
\subsection{Optimization}
We leverage the teacher-forcing technique to train the generative decoder with a cross-entropy loss given by
\begin{equation}
\begin{aligned}
\mathcal{L}_{ce}&= -\log {p(\bm{A}|(\bm{G},\bm{S},\bm{P}); \theta_{D_{g}})},\\
&= - \ {\sum \limits_{t=0}^{T_a} \log} p(\bm{a}_{t}|\bm{a}_{<t},\bm{F}_{fused}; \theta_{D_{g}}).
\end{aligned}
\vspace{-0.1cm}
\end{equation}
\ {Here we omit the average over the batch and time dimensions for simplicity. We follow this notation in the equations below if not otherwise specified.}
During inference, the actual predicted token is used as previous tokens frame by frame. To ensure a good speech intelligibility, we also introduce a \ {$L_1$-based} feature loss $\mathcal{L}_{feat}$ on the reconstructed waveform with the $11$-th layer feature of mHuBERT as target. Moreover, to enhance the semantic representation of $\bm{S}$, we employ a semantic distillation loss $\mathcal{L}_{se}$ on the quantized semantic token $\bm{S}$ after a linear projector, i.e. $\bm{S'}$, taking the $11$-th layer feature of mHuBERT as target. The total loss is a weighted combination of all terms as follows
\begin{equation}
\begin{aligned}
\mathcal{L} &= \mathcal{L}_{ce} + \lambda_{feat} \mathcal{L}_{feat} + \lambda_{se} \mathcal{L}_{se} + \lambda_r \mathcal{L}_{rate}.
\end{aligned}
\vspace{-0.2cm}
\end{equation}
\begin{equation}
\ {\mathcal{L}_{se} = -\sum \limits_{t=0}^{T-1} \log \sigma(cos(\bm{S'}^{(t,:)},\bm{S}_{se}^{(t,:)})).}
\vspace{-0.2cm}
\end{equation}
\ {Here $cos(\cdot)$ and $\sigma(\cdot)$ are cosine similarity and sigmoid activation, respectively. The superscript $(t,:)$ denotes the $t$-frame of the feature.} Except the pretrained local semantic encoder $E_s$ and acoustic decoder $D_a$, all other modules are optimized together in a single stage. \ {We set $\lambda_{feat}=1.0$, $\lambda_{se}=2.5$ and $\lambda_{r}=1.0$ to balance different terms in our experiments.}

\begin{table*}[!t]
%\fontsize{7}{7}\selectfont
\begin{center}
\caption{Evaluation of speech resynthesis quality. SPK-O: speaker similarity with original signal. Emotion: emotion accuracy evaluated on RAVDESS dataset. Other metrics are evaluated on LibriSpeech \textit{test-clean} for English and the Spanish test set of MLS for Spanish. We use \textbf{bold} to indicate the best result.}
\vspace{-0.1cm}
\begin{tabular}{l@{}c@{}c@{}ccccc@{\extracolsep{4pt}}c@{\extracolsep{4pt}}c@{\extracolsep{4pt}}c@{\extracolsep{4pt}}c}
\toprule
\textbf{Dataset} & \textbf{Methods} & \textbf{3s prompt} &\textbf{ Bitrate} & \textbf{Frame Rate}  & \textbf{WER}$\downarrow$ & $\bm{F}_{0}$\textbf{(VDE/FFE)}$\downarrow$ & \textbf{SPK-O}$\uparrow$ & \textbf{Emotion}$\uparrow$ & \textbf{NISQA}$\uparrow$ & \ {\textbf{VISQOL}$\uparrow$} & \ {\textbf{UTMOS}$\uparrow$}\\
\midrule
\multirow{10}{*}{English} 
% \rowcolor{gray!20}
& \cellcolor{gray!20} GT & \cellcolor{gray!20}$\times$ &\cellcolor{gray!20} - &\cellcolor{gray!20} 16 KHz  &\cellcolor{gray!20} 2.04 &\cellcolor{gray!20} 0.00/0.00 &\cellcolor{gray!20} 1.00 &\cellcolor{gray!20} 0.94 &\cellcolor{gray!20} 3.89&\cellcolor{gray!20} \ {5.00} &\cellcolor{gray!20}\ {4.10}\\
& EnCodec & $\times$ & 6 kbps & 75 Hz  & 2.20 & 8.60/8.67 & 0.90 & 0.88& 3.16&\ {4.27}&\ {3.10}\\
& TF-Codec & $\times$ & 6 kbps & 50 Hz & 2.16 & 2.85/2.88& 0.93 & 0.92& \textbf{3.87} &\ {4.50}&\ {3.92}\\
& SpeechTokenizer & $\times$ & 6 kbps& 75 Hz & 2.33& 5.56/5.64&0.85 & 0.89& 3.74&\ {4.27}&\ {3.92}\\
&  \ {DAC} & $\times$ & 6 kbps&  75 Hz &\textbf{2.09}& \textbf{2.81/2.85}&\textbf{0.95} & \textbf{0.93}& 3.80&\ {\textbf{4.69}}& \ {\textbf{4.02}}\\
\cmidrule(lr){2-12}
& \ {EnCodec-stg1} & $\times$ & 750 bps& 75 Hz  &39.58 & 18.68/19.61& 0.26 &0.56& 1.48 &\ {3.00}&\ {1.42}\\
& SpeechTokenizer-stg1 & $\times$ & 750 bps& 75 Hz &6.67 & 24.63/24.96& 0.18 &0.35& 1.11 &\ {2.13}&\ {1.26}\\
%& mHuBERT-Vocoder  & 500 bps & 50 Hz  & 0.0786 & 20.28/28.34 & 0.5682 & 0.6937 & 3.3441\\
& Hierarchical LM & $\surd$ & 500 bps & 50 Hz & 4.04 & 13.47/16.02 & 0.65 & 0.72& 3.85 &\ {2.98}&\ {3.38} \\
& UniCodec & $\times$ & 500 bps & 25 Hz & 3.03& 8.98/9.12& 0.71 & 0.79 & \textbf{3.94}&\ {3.53}&\ {3.56} \\
& UniCodec w. prompt &$\surd$ & 500 bps & 25 Hz & \textbf{2.93}&\textbf{8.45/8.56} & \textbf{0.76} & \textbf{0.83}& 3.92&\ {\textbf{3.63}}&\ {\textbf{3.61}}\\
\midrule
\multirow{10}{*}{Spanish} 
% \rowcolor{gray!20}
&\cellcolor{gray!20} GT &\cellcolor{gray!20} $\times$ &\cellcolor{gray!20} - &\cellcolor{gray!20} 16 KHz  &\cellcolor{gray!20} 12.00 &\cellcolor{gray!20} 0.00/0.00 &\cellcolor{gray!20} 1.00&\cellcolor{gray!20} -&\cellcolor{gray!20} 3.72 &\cellcolor{gray!20}\ {5.00}&\cellcolor{gray!20}\ {2.99}\\
& EnCodec & $\times$ & 6 kbps & 75 Hz  & 14.10 & 9.81/9.98 & 0.91& - & 2.92&\ {4.23}&\ {2.06}\\
& TF-Codec & $\times$ & 6 kbps & 50 Hz  & \textbf{12.74 }& 3.61/3.68 & 0.94 & -& \textbf{3.51}&\ {4.48}&\ {2.74}\\
& SpeechTokenizer & $\times$ & 6 kbps& 75 Hz & 17.24& 6.65/6.80& 0.84&- & 3.44&\ {4.15}&\ {2.63}\\
&\ { DAC} & $\times$ & 6 kbps& 75 Hz &12.87& \textbf{3.18/3.27}&\textbf{0.97} & -& 3.49&\ {\textbf{4.68}}&\ {\textbf{2.80}}\\
\cmidrule(lr){2-12}
& \ {EnCodec-stg1} & $\times$ & 750 bps& 75 Hz  &75.27 & 22.70/24.28& 0.27 &-& 1.45 &\ {2.92}&\ {1.23}\\
& SpeechTokenizer-stg1 & $\times$ & 750 bps& 75 Hz & 61.95& 27.80/27.98& 0.11 &- & 1.11&\ {1.96}&\ {1.24}\\
%& mHuBERT-Vocoder  &  500 bps & 50 Hz  & 0.1675 & 20.16/31.69& 0.6002 & 0.7502 & 3.4126 \\
& Hierarchical LM & $\surd$ & 500 bps & 50 Hz & 18.59& 14.55/17.66&  0.66&-& 3.46&\ {2.78}&\ {2.17}   \\
& UniCodec & $\times$ & 500 bps & 25 Hz & 17.13& 9.80/10.09& 0.77&-& \textbf{3.50} &\ {3.48}&\ {2.43}  \\
& UniCodec w. prompt &$\surd$ & 500 bps & 25 Hz & \textbf{16.66}&\textbf{9.46/9.72}&   \textbf{0.79}&-& 3.49&\ {\textbf{3.54}}&\ {\textbf{2.47}} \\
\bottomrule
\end{tabular}
\label{Tab_resyn}
\end{center}
%\vspace{-0.2cm}
\end{table*}

\section{UniCodec for Downstream Tasks}
Building upon the learned UniCodec tokens, we construct the UniCodeLM framework to facilitate prosody-aware speech language modeling. We design zero-shot text-to-speech synthesis (TTS) and speech-to-speech translation (S2ST) to explore the incorporation of paralinguistic information into speech language models, showing the effectiveness of our unified tokens in the generation framework. Moreover, for a comprehensive understanding of the information within our tokens, we also preform automatic speech recognition (ASR) for semantic evaluation and speech emotion recognition (SER) for paralinguistic evaluation.

% To show the capability of our UniCodec tokens, we designed several typical downstream tasks for evaluation, automatic speech recognition (ASR) for semantic evaluation, direct speech-to-speech translation (S2ST) for both understanding and generation, speech emotion recognition (SER) for paralinguistic representation evaluation, and zero-shot voice conversion (VC) for disentanglement evaluation. In downstream tasks, the pretrained UniCodec is freezed and we extract unified tokens from this codec to be fed into each downstream model. 

\vspace{-0.4cm}
\subsection{Zero-shot Text-to-Speech Synthesis (TTS)} 
Zero-shot TTS aims to synthesize speech given text for any target speaker. It is a typical one-to-many mapping problem, where identical text inputs yield varied speech sequences owing to diverse speech attributes like prosody, duration and volume variations. A recent neural codec language modeling TTS approach, VALL-E\cite{valle}, follows an acoustic modeling paradigm that directly maps the input text to acoustic tokens. However, due to the uncertainty of the target, such one-stage mapping approach often suffers from robustness issues \ {as reported in \cite{valle, ralle}}. As our UniCodec tokens are much more compact and semantic-rich than acoustic tokens, it reduces such uncertainty with better robustness. We predict the triplet tokens by 
%Benefiting from our token-based generation framework, we decompose the text-to-acoustic mapping into two stages: text-to-semantic/residual tokens and semantic/residual tokens-to-acoustic tokens, given by
\begin{equation} 
\begin{aligned}
& (\hat{\bm{S}},\hat{\bm{P}})=UniCodecLM(text;\tilde{x}), \hat{\bm{G}}=E_{g}(\tilde{x}),\\
\end{aligned}
\label{Eq_tts2}
\end{equation}
where $\tilde{x}$ is a speech prompt and $text$ is the text input. The $UniCodecLM$ is an autoregressive codec language model conditioned on $text$ and $\tilde{x}$, formulated as 
\begin{equation} 
\begin{aligned}
\mathcal{L}_{TTS}=-\ {\sum \limits_{t=0}^N \log} 
 p(\bm{c}_{t}|\bm{c}_{<t},\bm{Z}_{phoneme};\theta_{TTS}),
\end{aligned}
\vspace{-0.2cm}
\label{Eq_tts}
\end{equation}
where $\bm{c}_t=[\bm{s}_t,\bm{p}_t]$ denotes the concatenated semantic tokens $\bm{s}_t$ and residual tokens $\bm{p}_t$ of the $t$-th frame in $\bm{S}$ and $\bm{P}$ along the channel dimension. \ {$N$ denotes the length of UniCodec token sequence, and $\bm{c}_{-1}$ and $\bm{c}_{N}$ denote the $start$ and $eos$ tokens, respectively.} $\bm{Z}_{phoneme}$ is the phoneme transcription of the input text. We use a decoder-only transformer with 12 layers, 16 heads 
and a dimension of 768 for $\theta_{TTS}$. For better speech intelligibility of the synthesized speech,  we assign higher weights to the loss associated with $\bm{s}_t$ to direct the model's focus more on predicting semantic information. With predicted triplet token from Eq. (\ref{Eq_tts2}), the final speech is synthesized by the UniCodec decoder give by $\hat{x}=UniCodecDec(\hat{\bm{G}},\hat{\bm{S}},\hat{\bm{P}})$.  
% formulated as:
% \begin{equation}
% \bm{G} = E_{g}(\tilde{x}),
% \hspace{0.1cm}
% \hat{x} = {UniCodecDec} {(\bm{G},\bm{\hat{S}},\bm{\hat{P}})}.
% \end{equation}
%Notably, the second stage utilized the pretrained unicodec decoder, eliminating the need for retraining. 

\textbf{\textit{Connection with Two-Stage Approaches}} It is noted that although our approach takes one-stage modeling, it shares some connections with two-stage approaches \ {\cite{SPEAR-TTS} under} the Hierarchical LM paradigm, where semantic tokens \ {are used} as the intermediate representation between text and acoustic tokens. Compared with these methods, our \ {UniCodecLM} takes both semantic and residue tokens as the intermediate representation \ {if we take the generative decoder of UniCodec as the second stage.} It has several advantages: 1) the prosodic prediction is moved to the first stage and the global information is explicitly extracted from the prompt, alleviating the burden on acoustic generation. Experimental results show that this approach can significantly improve the stability and robustness of acoustic generation. 2) the UniCodec codes $(\bm{G},\bm{S},\bm{P})$ are jointly optimized with the generative decoder, ensuring a high-quality acoustic generation from the intermediate representations. 3) the modeling of both semantic and prosodic information in the first stage \ {within} a compact token \ {space} makes the prosodic prediction easier, leading to better expressiveness of the synthesized speech.
% 2) the semantic and prosody tokens were joint optimized with the generative decoder in the token learning paradigm, which makes the acoustic generation easier.

\vspace{-0.3cm}
\subsection{Speech-to-Speech Translation (S2ST)} 
S2ST aims to directly translate speech from the source language to the target with not only linguistic accuracy but also natural target-language output quality with all paralinguistic information preserved. Recently, unit-based methods have become more popular \cite{DS2ST,UnitY}, which typically follow a speech-to-unit pipeline with target speech tokens obtained from HuBERT \cite{HuBERT} as training target. As the target unit typically discards paralinguistic information, they can generate natural-sounding output with a target-language unit vocoder but the paralinguistic information of the speaker is not preserved, such as emotion, tone, intonation, speaking styles, etc. Additionally, a separate vocoder needs to be trained for each language.  As our UniCodec encapsolutes all linguistic and paralinguistic information into the tokens, we can achieve more natural target-language output. It is formulated as
\begin{equation} 
\begin{aligned}
(\hat{\bm{S}}_{tgt},\hat{\bm{P}}_{tgt})=UniCodecLM(\bm{S}_{src},\bm{P}_{src}), \hat{\bm{G}}_{tgt}=\bm{G}_{src},\\
\end{aligned}
\label{Eq_s2st}
\end{equation}
where the $UniCodecLM$ translates not only the linguistic part, but also the prosody from source language/content to the target domain simultaneously, therefore well preserves the speaking attributes. 
% Different from previous approaches, our objective is to evaluate the capability of the discrete tokens, thus we use a standard encoder-decoder transformer $\theta_{S2ST}$ as our downstream model with a unit-to-unit translation pipeline, similar to \cite{U2UT}.
Specifically, we concatenate our semantic and residue tokens, $\bm{S}$ and $\bm{P}$, as the input and target units and the training objective is given by 
\begin{equation} 
\begin{aligned}
\hspace{-1mm}\mathcal{L}_{S2ST}\hspace{-1mm}=\hspace{-1mm}-\hspace{-1mm} \ {\sum \limits_{t=0}^N \log}  p(\bm{c}_{tgt,t}|\bm{c}_{tgt,{<t}},\bm{C}_{src},\bm{L}_{src},\bm{L}_{tgt};\theta_{S2ST}),
\end{aligned}
\end{equation}
where $\bm{C}=[\bm{S},\bm{P}]$ denotes the concatenated semantic tokens $\bm{S}$ and residual tokens $\bm{P}$ along the channel dimension and $\bm{c}_t$ is the $t$-th frame token of $\bm{C}$, a similar setting as that in TTS. $\bm{L}_{src}$ and $\bm{L}_{tgt}$ denote the source and target language tokens.
We use an encoder-decoder transformer with a 12-layer encoder and a 12-layer decoder with 8 heads and a dimension of 512 for $\theta_{S2ST}$. The translated output $\bm{\hat{C}}_{tgt}=(\hat{\bm{S}}_{tgt},\hat{\bm{P}}_{tgt})$ will be combined with the global tokens from source $\hat{\bm{G}}_{tgt}=\bm{G}_{src}$ to synthesis the target speech through the UniCodec decoder by $\hat{x}=UniCodecDec(\hat{\bm{G}}_{tgt},\hat{\bm{S}}_{tgt},\hat{\bm{P}}_{tgt})$.
% \begin{equation}
% \hat{x} &= {UniCodecDec} {(\bm{G}_{src}},\bm{\hat{S}_{tgt}},\bm{\hat{P}_{tgt}})}.
% \end{equation}

% \textbf{Zero-Shot Voice Conversion (VC)} Here we focus on zero-shot voice conversion. As our UniCodec tokens are disentangled, we can directly use the learned codec for voice conversion without extra training. Specifically, we simply replace the global token $\bm{G}$ of the source speech with the target one, keep the source $\bm{S}$ and $\bm{P}$, and synthesize using the codec decoder, i.e.
% \begin{equation}
% \hat{x} &= {UniCodecDec} {(\bm{G}_{tgt},\bm{S}_{src},\bm{P}_{src})}.
% \end{equation}

\vspace{-0.2cm}
\subsection{Speech and Emotion Recognition\label{downstream_C}} 
\textbf{Automatic Speech Recognition (ASR)} ASR aims to transcribe speech into text content. We take the discrete tokens extracted from the pretrained UniCodec as input and leverage a decoder-only language model to predict text tokens auto-regressively to evaluate the semantic representation capability.
%There are two experiment settings for our downstream ASR model. The first takes a vanilla 2-layer 1024-unit BLSTM as the downstream head. Specifically, we take the discrete tokens extracted by the pretrained codec as input and optimized the downstream head by CTC loss on characters. No languange model is used during decoding.
% is the SUPERB\cite{SUPERB} evaluation benchmark. Notably, in stead of taking the weighted average of each pretrained model layer, we use the discrete token embedding as the input 
%The second setting aims to emulate the setting of LLM, where we measure the amount of the semantic information captured by the tokens through a decoder-only language model. 
Specifically, we adopt a 12-layer decoder-only transformer with 8 heads as the downstream ASR model, optimized by 
\begin{equation}
%\begin{aligned}
\mathcal{L}_{ASR}=-\ {\sum \limits_{i=0}^N \log}  p(\bm{z}_{i}|\bm{z}_{<i},[\bm{S},\bm{P}];\theta_{ASR}),
%\end{aligned}
\end{equation}
where $\bm{z}$ denotes the text token sequence \ {and $\bm{z}_{i}$ is the $i$-th token of $\bm{z}$}. 
% We use the pretrained \texttt{Wav2Vec2CTCTokenizer}\footnote{\footnotesize \url{https://huggingface.co/jonatasgrosman/wav2vec2-large-xlsr-53-english}} as our text tokenizer and the vocabulary size is 33. 

\textbf{Speech Emotion Recognition (SER)} SER aims to predict a global emotion class $l_{emotion}$ from the speech. We choose SER to evaluate the emotional representation capability and paralinguistic information captured by our tokens. We first use a temporal pooling layer to aggregate $\bm{S}$ and $\bm{P}$ into two global embeddings, concatenate them with $\bm{G}$, and adopt a linear classifier as the downstream model, which is formulated as
\begin{equation}
%\begin{aligned}
\mathcal{L}_{SER}=-\log p(l_{emotion}|[\bm{G}, Pooling(\bm{S},\bm{P})];\theta_{SER}).
%\end{aligned}
\end{equation}

% two are among the most widely used datasets. 
% We found that in IEMOCAP there are instances of background noise such as laughter and also instances where multiple people are speaking simultaneously, which significantly impacts the effectiveness of the codec, other works also mentioned that the codec-related models are sensitive to the background environment \cite{}.
% Therefore, we use RAVDESS and CREMA-D for SER.
% RAVDESS is a multi-modal database of emotional speech and song with 8 different emotions: happy, sad, angry, fearful, surprise, disgust, calm, and neutral. We merged the neutral and calm emotions, happy and surprised emotions, resulting in 6 emotions, and used the first 20 actors for training, actors 20-22 for validation and 22-24 for test. 
% For CREMA-D, we split the dataset into 70\% for training, 10\% for validation and 20\% for testing, each with no overlapped speakers.

%\begin{figure*}[ht] 
%\vspace{-0.3cm}
%\centering 
% \includegraphics[width=0.85\linewidth]{figure/emotion_tsne.png}
%\vspace{-0.3cm}
%\caption{T-SNE of emotion information preserved in the reconstructed speech on RAVDESS dataset.}
%\vspace{-0.5cm}
%\label{Fig_emotion} 
%\end{figure*}

\vspace{-0.2cm}
\section{Experimental Results}
In the experiments, we evaluate our UniCodec from several aspects.
% including speech synthesis quality from discrete tokens, capability on various understanding and generation downstream tasks, and the disentanglement of the triplet tokens. 
We first conduct speech resynthesis from discrete tokens to assess the preservation of various information of speech, including content, timbre, prosody, emotion in Section \ref{exp_A}, and then investigate the disentanglement of the triplet UniCodec tokens in Section \ref{exp_B}. Section \ref{exp_C} evaluates the tokens' representation capability by ASR for semantic understanding and SER for paralinguisitic understanding. Section \ref{exp_D} conducts a deep analysis on speech generation using our UniCodec tokens by TTS and S2ST. \ {All audios used in our study are resampled to 16 kHz.}

\subsection{Speech Resynthesis\label{exp_A}}
The UniCodec is trained on a 6k-hour multilingual dataset, a combination of the Libri-Light medium subset \cite{Librilight}, the Spanish, French, Mandarin, German, Italian, and Russian subsets of the ICASSP 2021 DNS Challenge corpus \cite{DNSChallenge}, and AVSpeech \cite{AVSpeech}. Each utterance is cut into 3 to 20 seconds \ {with a variable length}. We use the \textit{dev-clean} subset of LibriSpeech for validation, the \textit{test-clean} subset of LibriSpeech and the Spanish test set of MLS \cite{MLS} for testing. Besides, we also use another emotional speech dataset RAVDESS\cite{RAVDESS} to evaluate the emotion preservation in the synthesized speech. 

In UniCodec, we set the target bitrate of $\bm{S}$ and $\bm{P}$ tokens to be 250 bps, respectively, leading to a total of 500 bps. \ {The group number $M$, the number of codewords $K$ in GVQ are set to 2 and 256, respectively, with both $R^s_{target}$ and $R^p_{target}$ set to 10 bits per $40ms$ frame.} For global tokens $\bm{G}$, we directly use the continuous embedding \ {without a quantization as it can be extracted from speech prompt in downstream tasks as described in Section IV, without a need to generate it. As global embedding typically takes a very small portion of the total bitrate in coding, considering its consistency over the long stretches of speech \cite{lowbitrate}, we only count the bits for $\bm{S}$ and $\bm{P}$ in comparison with other acoustic codecs here.}  \ {The channel dimensions $C_{g}$, $C_{a}$, $C_{s}$, $C_{p}$ are 256, 768$\times$2, 1024 and 1024, respectively. We use a 12-layer transformer decoder for the generative decoder module, with a hidden dimension of 1024, a feed-forward dimension of 4096, and 8 attention heads. 
%The STFT in $E_g$ uses a Hanning window with a window length of 40 ms and a hop length of 10 ms.
} We use adam optimizer with an initial learning rate of 4e-4 for training. The batch size is set to 40 seconds per GPU \ {and the UniCodec is trained on 16 V100 GPUs}. 

% We use speech resynthesis from discrete tokens to assess the preservation of various aspects of speech information, including content, timbre, prosody, emotion, et al. 
We compare with several baselines:
%\textbf{mHuBERT-Vocoder}, which uses the publicly released mHuBERT $11$-th-layer unit after k-means as the semantic token and a HiFi-GAN\cite{hifigan} vocoder to perform a unit-to-waveform mapping. This vocoder is language-specific and trained on single-speaker datasets.

\textbf{Acoustic Codec}, which includes \ {three state-of-the-art high-quality} neural speech codecs, EnCodec \cite{EnCodec}, TF-Codec \cite{TF-Codec} and \ {DAC \cite{dac}}. They encode speech into fine-grained acoustic tokens and produce high-fidelity reconstruction through their acoustic decoders.  

\textbf{SpeechTokenizer \cite{SpeechTokenizer}}, which is similar to EnCodec but it encodes semantic information in the first stage of RVQ and other detailed information in subsequent stages.

\textbf{Hierarchical LM}, which uses the publicly released mHuBERT $11$-th layer unit after k-means clustering as the semantic tokens and a semantic-to-acoustic mapping model to generate TF-Codec acoustic tokens autoregressively for speech resynthesis, with a $3$-\textit{second} speech prompt randomly cropped from the test speech. \ {We only count the bitrate for the semantic tokens without calculating that for the prompt as it is not a neural codec. We add it just to show the difference with UniCodec in information recovery from available tokens.}
% For a fair comparison, the Hierarchical LM uses the same acoustic token and acoustic decoder as the UniCodec and the same downstream model is used for all baselines. 
% todo : baseline描述

%For evaluating the discrete tokens, we compare their bitrate and the capability of preserving speech information.
To evaluate the reconstructed quality, we leverage several metrics that focus on different aspects, including:
(i) \ {Word error rate (WER)}, which measures the speech intelligibility. we use two public ASR models for English\footnote{  \url{https://huggingface.co/facebook/hubert-large-ls960-ft}} and Spanish \footnote{\url{https://huggingface.co/jonatasgrosman/wav2vec2-large-xlsr-53-spanish}}, respectively. (ii) \ {Voicing Decision Error (VDE) \cite{VDE} and $F_0$ Frame Error (FFE) \cite{FFE}}, $F_{0}$-related metrics which assess the accuracy of voicing decisions and pitch values. We use them to evaluate the preservation of prosodic information. (iii) SPK-O, cosine similarity of speaker embeddings between the synthesized and the original speech, extracted by a speaker verification model WavLM-TDNN\footnote{  \url{https://github.com/microsoft/UniSpeech/tree/main/downstreams/speaker_verification}}. (iv) Emotion-Acc, the top-1 emotion accuracy calculated by an emotion recognition model\footnote{\url{https://huggingface.co/ehcalabres/wav2vec2-lg-xlsr-en-speech-emotion-recognition}} on RAVDESS test set.
(v) \ {The commonly used NISQA \cite{NISQA}, UTMOS \cite{utmos} and VISQOL \cite{vsiqol} metrics, which evaluate the naturalness and the overall quality of the synthesized speech. Among them, except NISQA and UTMOS, all others like ASR-WER, VDE, FFE, SPK-O, Emotion-Acc, and VISQOL are metrics with a reference to measure the preservation of various speech information, i.e. the fidelity of reconstructed speech, an indispensable goal for low-bitrate and generative codecs.}

Table \ref{Tab_resyn} summarizes the speech reconstruction results. We could see that when operating at high bitrates, all the three acoustic codecs and SpeechTokenizer could deliver relatively high-fidelity audio, indicating that the fine-grained acoustic tokens could preserve all aspects of information effectively. \ {Among them, DAC demonstrates the highest overall quality, underscoring its superiority. It is noted that DAC is a non-causal and long-latency codec while the other acoustic codecs are all causal and low-latency systems. The TF-Codec used as acoustic target in our UniCodec is still on par with DAC despite its causal and low-latency property.}
% The mHuBERT-Vocoder baseline, that solely uses a unit-vocoder to map mHuBERT units back to waveforms, struggles to ensure the preservation of intonation and rhythm in the generated audio, resulting in very low scores for F0, speaker and emotion similarity.

\begin{figure}[tb] 
\centering 
\vspace{-0.2cm}
\includegraphics[width=0.8\linewidth]{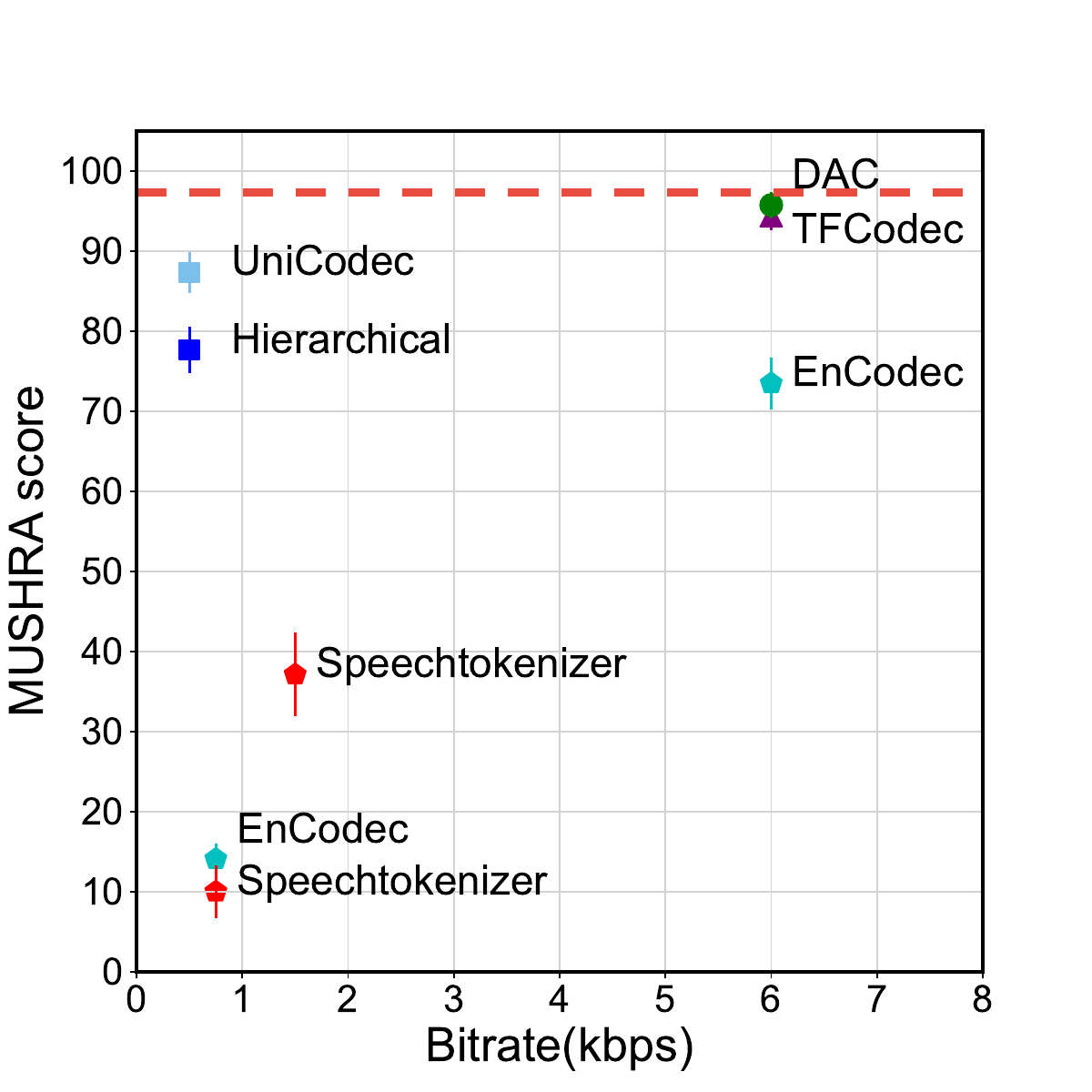}
\caption{\ {Subjective evaluation results. The red dotted line represents the score of the reference. The error bar denotes $95\%$ confidence intervals.}}
\vspace{-0.4cm}
\label{Fig_mushra} 
\end{figure}

When only the semantic tokens of SpeechTokenizer is used for speech resynthesis (denoted as SpeechTokenizer-stg1 in Table \ref{Tab_resyn}), \ {it achieves good content preservation with a much lower WER compared to EnCodec using only the first stage of RVQ (denoted as EnCodec-stg1). However, a significant loss on paralinguistic information is observed, indicated by all other metrics.} Compared to SpeechTokenizer-stg1, Hierarchical LM achieves better retention of timbre and emotion information, benefiting from the in-context learning from the $3$-\textit{second} speech prompt; however, there is still a considerable loss in $F_{0}$ \ {as the prosody generation such as intonation has large diversities by in-context learning from a speech prompt}. Our UniCodec, utilizing only 500 bps tokens, outperforms \ {all three low-bitrate baselines} and approaches high-bitrate acoustic codecs in terms of naturalness, intelligibility and preservation of paralinguistic information and acoustic details, demonstrating the high expressiveness of our tokens and the high reconstruction quality of UniCodec. \ {What's more, despite a generative codec, the UniCodec maintains a good speech fidelity, keeping both linguistic and paralinguistic information well.} The higher scores in $F_{0}$, emotion, and timbre compared to the Hierarchical LM indicate that relying solely on in-context learning to obtain paralinguistic information from prompts is insufficient. Our UniCodec explicitly models paralinguistic information, facilitating better preservation of speech. 

\ {To further assess the audio reconstruction quality, we also conduct a subjective listening test by the MUSHRA-inspired
crowd-sourced method \cite{mushra}, where 10 participants evaluate 10 samples randomly sampled from the LibriSpeech \textit{test-clean} test set. Each listener is presented with a hidden reference and a set of test samples generated by different methods. As shown in Fig. \ref{Fig_mushra}, we can observe that under low bitrates (lower than 1.5 kbps), the audio quality reconstructed by EnCodec and SpeechTokenizer significantly degrades; while our UniCodec, operating as an ultra-low bitrate codec, outperforms all the low-bitrate baselines and performs close to high bitrate acoustic codecs DAC and TF-Codec, demonstrating its superior perceptual quality.}

It should also be noted that although in speech resynthesis, our UniCodec is slightly poorer than high-bitrate neural codecs, our tokens are more compact and easy to predict in speech language models, showing much better quality in downstream tasks such as TTS and S2ST as shown in Section \ref{exp_D}. \ {Besides, although we leverage only TF-codec as the acoustic target in UniCodec, the proposed framework of learning the unified token for speech language modeling is applicable to any other high-quality codecs as well with corresponding acoustic generation.}
%%todo 加上prompt更好
% Fig. \ref{Fig_emotion} shows the T-SNE illustration of the emotion embeddings extracted from different synthesized output, which also verifies the good preservance of emotion information in UniCodec. 
% We put some samples on our demo page. 
% \footnote{ \url{https://unicodec.github.io/demo/}}
% %%todo: multilingual 能力

%https://www.mdpi.com/1424-8220/21/4/1249
\begin{table}[!t]
\begin{center}
\caption{Evaluation of token disentanglement by zero-shot voice conversion. \ {SPK-O: speaker similarity with the original signal.}}
\begin{tabular}{lc@{\extracolsep{4pt}}c@{\extracolsep{4pt}}c@{\extracolsep{4pt}}c@{\extracolsep{4pt}}c}
\toprule
%\hline
Methods & SPK-O$\uparrow$ & WER$\downarrow$ & NISQA$\uparrow$  & \ {MOS-N$\uparrow$}& \ {MOS-S$\uparrow$}  \\ 
\midrule
%\hline
YourTTS-VC  & 0.42 & 4.59 & 3.44 &\ {2.46} &\ {2.51}\\
Hierarchical LM  & 0.59 & 4.28 & 3.91&\ {3.86}&\ {3.91}  \\
UniCodec & 0.55  & 2.88 & \textbf{3.99}&\ {4.02}&\ {4.11} \\ 
%UniCodec & 0.8405  & 0.0287 & 3.9768 \\ 
UniCodec w. prompt &  \textbf{0.60} & \textbf{2.84} & 3.93&\ {\textbf{4.10}}&\ {\textbf{4.22}} \\ 
%\hline
\bottomrule
\end{tabular}
\label{Tab_VC}
\end{center}
\vspace{-0.3cm}
\end{table}

%\begin{table}[!t]
%\vspace{-0.2cm}
%\begin{center}
%\caption{Ablation study on emotion recognition.}
%% \vspace{0.2cm}
%\begin{tabular}{lc}
%\toprule
%Methods & Accuracy\uparrow\\
%\midrule
%Global  &  52.92 \\
%Semantic &  43.93  \\
%Semantic+residue &  50.42\\
%Global+semantic &  55.94  \\
%Global+semantic+residue & \textbf{57.93}  \\
%\bottomrule
%\end{tabular}
%\label{Tab_emotion}
%\end{center}
%\vspace{-0.3cm}
%\end{table}

% \begin{table}[!t]
% %\fontsize{7}{7}\selectfont
% \begin{center}
% \caption{Subjective evaluation of zero-shot text-to-speech synthesis. * denotes that the model is trained on LibriTTS dataset.
% \label{Tab_tts2}}
% \vspace{-0.1cm}
% \begin{tabular}{lcc}
% \toprule
% Methods & CMOS$\uparrow$ & SMOS$\uparrow$ \\
% \midrule
% Groundtruth &  &    \\
% YourTTS &  &    \\
% VALL-E* &  &    \\
% SpeechTokenizer* &  &    \\
% \midrule
% Hierarchical LM* &  &    \\
% UniCodec* &  &    \\
% UniCodec &  &    \\
% % \multirow{4}{*}{English} 
% % & Groundtruth & &  &  &  \\
% % & TF-Codec &  &  &  &  \\
% % & Hierarchical LM & &  &  & \\
% % & UniCodec &  &  &  &  \\
% \bottomrule
% \end{tabular}
% \end{center}
% %\vspace{-0.2cm}
% \end{table}

\vspace{-0.3cm}
\subsection{Feature Disentanglement\label{exp_B}}
In this subsection, we conduct two additional experiments, voice conversion and pitch conversion, to evaluate the disentanglement of the triplet tokens $\bm{S},\bm{P}$ and $\bm{G}$ in our UniCodec.

\textbf{Voice Conversion} To show the decoupling of speaker identity and content information, we perform zero-short voice conversion by simply replacing the global token $\bm{G}$ of the source speech with the target one, keeping the source $\bm{S}$ and $\bm{P}$, and synthesizing target speech using the UniCodec decoder, given by $\hat{x} = {UniCodecDec}{(\bm{G}_{tgt},\bm{S}_{src},\bm{P}_{src})}$. For evaluation, we randomly sample 10 speakers from LibriSpeech \textit{test-clean} subset, each with 3 utterances, and voice conversion is conducted between different speakers, resulting in 810 cases. We compare our UniCodec with YourTTS-VC \cite{yourtts} and the Hierarchical LM baseline that takes the target prompt to recover the speaker information. 
\ {We leverage both the objective metrics SPK-O, WER and NISQA as in the previous subsection, and the subjective mean opinion score (MOS) by human rating for evaluation, as shown in Table \ref{Tab_VC}. In subjective testing, 10 participants are invited to assess the naturalness of the synthesized speech (MOS-N) and the speaker/style similarity to the source (MOS-S) on 8 samples randomly drawn from the test set without source speaker overlapping. Both MOS-N and MOS-S span from 1 to 5, with higher values signifying better speech quality and higher speaker similarity, respectively.}

As shown in Table \ref{Tab_VC}, our UniCodec largely outperforms YourTTS-VC, demonstrating its good disentanglement of $\bm{G}$ with $\bm{S}$ and $\bm{P}$ and the high-quality generation. Compared with Hierarchical LM, our UniCodec achieves better intelligibility and overall quality due to better speech modeling. It can also be seen that the Hierarchical LM achieves a good speaker similarity, showing that speaker timbre could be efficiently transferred through in-context learning from the prompt. When we also use a $3$-\textit{second} voice prompt in our generative decoder, even better speaker similarity is achieved, shown as `UniCodec w. prompt'.

% Hierarchical LM also achieves better speaker similarity than YourTTS-VC, showing that speaker timbre could be efficiently transferred through in-context learning from the prompt. Compared with Hierarchical LM, our UniCodec achieves higher speaker similarity, better intelligibility and overall quality due to explicitly modeling of globel speaker information and better disentangled speech modeling. 

% Compared with Hierarchical LM, our UniCodec achieves better intelligibility and overall quality due to better speech modeling. It can also be seen that the Hierarchical LM achieves slightly better speaker similarity than UniCodec, showing that speaker timbre could be efficiently transferred through in-context learning from the prompt. When we also use the voice prompt in our generative decoder, even better speaker similarity is achieved, shown as 'UniCodec w. prompt'.
\begin{table}[!t]
\begin{center}
\caption{Evaluation of token disentanglement by pitch conversion. O: original signal, N: $F_{0}$-normalized signal. VDE/FFE-N means the metrics computed between $F_{0}$-normalized and converted signals. VDE/FFE-O means the metrics computed between the original and converted signals.  }
% \vspace{-0.1cm}
\begin{tabular}{lccc}
\toprule
Methods & Input (G/S/P) & VDE/FFE-N & VDE/FFE-O \\  %\hline
\midrule
\multirow{2}{*}{GT} & O/O/O   &  0.150 / 0.215 & 0.000 / 0.000\\ 
  & N/N/N   & 0.000 / 0.000 & 0.150 / 0.218 \\
\midrule
\multirow{2}{*}{UniCodec} 
  & O/N/O   & 0.150 / 0.215 & 0.070 / 0.071 \\
  & O/O/N   & 0.069 / 0.072 & 0.151 / 0.218 \\ 
\midrule
\multirow{2}{*}{UniCodec-w/o disen}   
 & O/N/O   &  0.109/ 0.139 & 0.100 / 0.122\\
 & O/O/N   &  0.125/ 0.172 & 0.082 / 0.092\\
\bottomrule
\end{tabular}
\label{Tab_F0}
\end{center}
\vspace{-0.3cm}
\end{table}

\textbf{Pitch Conversion} Here we conduct experiments to show the decoupling of semantic information in $\bm{S}$ and prosody information in $\bm{P}$. 
% As stated in many literature\cite{}, the prosody information contains several subcomponents: pitch, rhythm and tempo. 
We mainly focus on the pitch information, a major component of prosody that could be represented by fundamental frequency $F_{0}$, and conduct $F_{0}$ conversion for this disentanglement evaluation. Specifically, we randomly sample 100 utterances from LibriSpeech \textit{test-clean} subset. We normalize the $F_{0}$ of each utterance to a random value per 500 \textit{ms} segment. Pitch conversion is conducted between the original signal and the $F_{0}$-normalized target signal. In this experiment, we expect to achieve pitch conversion by replacing the residual tokens $\bm{P}$ with that of the target signal, given by $\hat{x} = {UniCodecDec}{(\bm{G}_{orig},\bm{S}_{orig},\bm{P}_{norm})}$. We evaluate the effectiveness of pitch conversion by computing VDE and FFE between the converted and the target signals. Lower VDE and FFE values indicate similar pitch characteristics. 

As shown in Table \ref{Tab_F0},  when we replace $\bm{P}$ with that from the $F_{0}$-normalized signal while keeping $\bm{S}$ and $\bm{G}$ unchanged (denoted as O/O/N), the synthesized signal shows very low VDE/FFE-N with the target signal. This suggests that the pitch is effectively transferred to that of the target signal. 
% We provide some examples on the demo page\footnote{Audio samples are available at \href{https://unicodec2.github.io/demo/}{https://unicodec2.github.io/demo/.}}. 
Besides, in a non-disentanglement variant of UniCodec shown as \ {UniCodec-w/o disen, which uses the same encoder for both $\bm{S}$ and $\bm{P}$ without any decoupling and discards the semantic loss $\mathcal{L}_{se}$,} neither replacing $\bm{S}$ nor $\bm{P}$ achieves satisfactory transfer effects. Additionally, we calculate the VDE/FFE-O between the converted and the original signals to assess the extent of disturbance to prosody information by modifying different tokens, shown as the last column in the Table \ref{Tab_F0}. When we only change $\bm{S}$ to that of the $F_{0}$-normalized signals (denoted as O/N/O), the $F_{0}$ contours of the synthesized signal are hardly affected with a relatively low VDE/FFE-O, which indicates that semantic token $\bm{S}$ contains minimal prosody-related information. In contrast, changing $\bm{P}$ to that of the $F_{0}$-normalized signal would largely disturb the pitch information, leading to a high VDE/FFE-O. For UniCodec-w/o disen, the prosody information is mixed with semantic information in both tokens without clear meaning. When we separately replace each token, the prosody information can still be partially restored from the other token, resulting in small pitch differences between the converted and the original signals.

%\textbf{Visualization}
%Fig. \ref{Fig_correlation} shows the correlation between the local semantic token $S$ and the residual one $P$. We concatenate $S$ and $P$, and calculate the covariance matrix. As both $S$ and $P$ use 2 groups for quantization (there are in total 4 groups), we can see that the correlation between $S$ and $P$ is low (bottom-left and top-right). Also the cross-group correlation is low, showing that in $S$ or $P$, the 2 groups are also de-correlated for efficient representation at such low bitrates. 
%\vspace{-0.1cm}
\begin{table}[!t]
\begin{center}
\caption{Evaluation of token representation on understanding tasks. ASR: automatic speech recognition; SER: speech emotion recognition.}
% \vspace{-0.1cm}
\begin{tabular}{lccc}
\toprule
%\hline
\multirow{2}{*}{Methods} & \multirow{2}{*}{ASR(WER)$\downarrow$} & \multicolumn{2}{c}{SER(Accuracy)$\uparrow$}   \\ 
 &   & RAVDESS & CREAMA-D  \\ 
\midrule
%\hline
TF-Codec & 40.31  & 40.00 &  48.17  \\ 
mHuBERT  & \textbf{9.33}  & 64.17 & 51.85  \\
%Hierarchical LM & - & - & -   \\
UniCodec & 9.44  & \textbf{73.33} & \textbf{57.93}  \\ 
%\hline
\bottomrule
\end{tabular}
\label{Tab_understanding}
\end{center}
\vspace{-0.3cm}
\end{table}

\vspace{-0.1cm}
\subsection{Token Representation Capability\label{exp_C}}
The above sections analyze what information has been encoded within tokens. In this section, we further explore the representative capabilities of these tokens for downstream speech understanding task, including ASR and SER, \ {as described in Section \ref{downstream_C}}.  

\textbf{Speech and Emotion Recognition}
\ {We conduct ASR experiments on the \textit{train-clean-100} subset of the LibriSpeech dataset\cite{librispeech} for training the downstream model and evaluate the WER on the \textit{test-clean} subset. 
 For SER, we conduct experiments on two widely used emotional speech datasets, RAVDESS \cite{RAVDESS} and CREMA-D \cite{CREMA-D}. For RAVDESS, 
 % we merge the neutral and calm emotions, happy and surprised emotions, resulting in 6 emotions, and
 we use the first 20 actors for training, actors 21-22 for validation and 23-24 for testing. 
For CREMA-D, we split the dataset with 70\% for training, 10\% for validation and 20\% for testing. The batch size is set to 32 for ASR and 16 for SER.} For ASR, a good token should be able to accurately represent the semantic information of speech and for SER, tokens should adequately capture the emotional and affective information present in speech, facilitating emotion classification or sentiment analysis. We compare our UniCodec tokens with acoustic tokens from TF-Codec and semantic tokens from mHuBERT.

As shown in Table \ref{Tab_understanding}, the TF-Codec achieves poor performance in both ASR and SER, showing that although all information are well preserved in acoustic tokens to deliver a high reconstruction quality, their representative capability or expressiveness on semantic information and emotion is poor. Comparable WER is achieved between our UniCodec and mHuBERT tokens, showing the adequate linguistic representation by our tokens through disentanglement. For emotion recognition, benefiting from better modeling of paralinguistic information in $\bm{G}$ and $\bm{P}$, our UniCodec tokens significantly outperforms mHuBERT, demonstrating the rich paralinguistic information within our tokens. In the next section, we will investigate how these paralinguistic information will benefit the speech language modeling in the generation framework.
% To examine the individual contributions of each token, we conducted ablation study. 

% As shown in Table \ref{Tab_emotion}, semantic tokens $\bm{S}$ carry little emotion information. With the inclusion of residual tokens, there is a significant increase in accuracy, indicating that the prosody information in residual tokens plays a crucial role for emotion recognition. There is a further big performance boost when combining them with global tokens as the global speaking style captured by $\bm{G}$ is also a big hint of emotion.

\begin{table*}[!t]
%\fontsize{7}{7}\selectfont
\begin{center}
\caption{Evaluation of zero-shot text-to-speech synthesis on LibriSpeech \textit{test-clean} with a duration between 4 and 10 seconds. $^\blacklozenge$ means the results are directly obtained from the paper. $^\lozenge$ means the results are obtained from the official released models. * means the results are obtained from third-party implementations since the official implementation is not publicly available.
SPK-D: speaker similarity with compressed prompt. SPK-O: speaker similarity with original prompt. LL-full: the full 60K-hour dataset of Libri-Light. LL-medium: the 6K-hour medium subset of Libri-Light. LT: LibriTTS training data. 
\label{Tab_tts}}
\vspace{-0.1cm}
\begin{tabular}{lccccc}
\toprule
\textbf{Methods} & \textbf{Training Data} & \textbf{NISQA$\uparrow$} & \textbf{WER$\downarrow$} &\textbf{ SPK-D$\uparrow$} & \textbf{SPK-O$\uparrow$}\\
\midrule
Groundtruth & - & 3.88 & 2.20 & -  & 0.74  \\
YourTTS$^{\lozenge}$ &  & 3.28& 7.95 & - & 0.47\\
VALL-E$^{\blacklozenge}$ & LL-full (60k hrs) & - & 5.90 & 0.58 & -\\
VALL-E$^{*}$ & LT (585 hrs)  & 3.31 & 18.67& 0.42 &0.34 \\
USLM$^{\blacklozenge}$  &  MLS (44k hrs)& - & 5.40 & \textbf{0.68} & -\\

USLM$^{\lozenge}$ &  LT (585 hrs) &3.39 & 18.66 & 0.40 &0.29\\
\midrule
Hierarchical LM &  LT (585 hrs) &3.92 & 4.76 $\pm$ 0.28 & 0.62 &0.53\\
UniCodecLM-w/o disen & LT (585 hrs) & 3.88& 5.81 $\pm$ 0.05 & 0.66&0.56\\
UniCodecLM-w/o weighted &  LT (585 hrs) &3.96 & 4.71 $\pm$ 0.11 & 0.66 &0.57 \\
UniCodecLM  &  LT (585 hrs) & \textbf{3.99} &\textbf{4.45} $\pm$ \textbf{0.03}  &  0.66&0.57 \\
\midrule
% TF-Codec & LL-medium (6k hrs) + LT (585 hrs) & & &  \\
Hierarchical LM &  LL-medium (6k hrs) + LT (585 hrs) &3.92 & 4.00 $\pm$ 0.30& 0.63 &  0.53\\
UniCodecLM-w/o disen & LL-medium (6k hrs) + LT (585 hrs)&3.87 & 4.90 $\pm$ 0.18 & 0.65 & 0.56\\
UniCodecLM-w/o weighted &  LL-medium (6k hrs) + LT (585 hrs) &3.98 & 3.69 $\pm$ 0.07 & 0.66 & 0.57 \\
UniCodecLM &  LL-medium (6k hrs) + LT (585 hrs) &\textbf{3.99} & \textbf{3.51 $\pm$ 0.03} & 0.66 & \textbf{0.57} \\
% \multirow{4}{*}{English} 
% & Groundtruth & &  &  &  \\
% & TF-Codec &  &  &  &  \\
% & Hierarchical LM & &  &  & \\
% & UniCodec &  &  &  &  \\
\bottomrule
\end{tabular}
\end{center}
%\vspace{-0.2cm}
\end{table*}

\begin{table}[!t]
%\fontsize{7}{7}\selectfont
\begin{center}
\caption{Evaluation of zero-shot text-to-speech synthesis on RAVDESS test set. SPK-D: speaker similarity with compressed prompt.}
\label{Tab_tts2}
\vspace{-0.1cm}
\begin{tabular}{lccc}
\toprule
\textbf{Methods }& \textbf{NISQA$\uparrow$} &\textbf{Emotion-Acc$\uparrow$} &\textbf{SPK-D$\uparrow$} \\
\midrule
Groundtruth & 3.11 & 0.94 & 1.00  \\
YourTTS$^{\lozenge}$ & 3.08 & 0.20 & 0.32\\
VALL-E* &  2.85 & 0.24 & 0.19\\
% USLM$^{\lozenge}$ & & & \\
% SpeechTokenizer
\midrule
% TF-Codec & & &  \\
Hierarchical LM & 3.12 & 0.38 & 0.46 \\
UniCodecLM-w/o disen & 3.23 & 0.47 &  0.51 \\
UniCodecLM-w/o weighted & 3.22 & 0.49 & 0.51 \\
UniCodecLM & \textbf{3.25} & $\textbf{0.49}$ & $\textbf{0.53}$ \\
% \multirow{4}{*}{English} 
% & Groundtruth & &  &  &  \\
% & TF-Codec &  &  &  &  \\
% & Hierarchical LM & &  &  & \\
% & UniCodec &  &  &  &  \\
\bottomrule
\end{tabular}
\end{center}
%\vspace{-0.2cm}
\end{table}

\begin{table}[!t]
%\fontsize{7}{7}\selectfont
\begin{center}
\caption{\ {Subjective evaluation of zero-shot text-to-speech synthesis on 12 samples randomly drawn from LibriSpeech \textit{test-clean} and RAVDESS test set. For fair comparison, both Hierarchical LM and UniCodecLM use the versions trained on LibriTTS with 585 hours in Table V. 
%MOS-N: Mean opinion score of naturalness. MOS-S on LibriSpeech: Mean opinion score of speaker voice similarity between the synthesized speech and the prompt. MOS-S on RAVDESS: Mean opinion score of speaking style similarity between the synthesized speech and the prompt.
}
\label{Tab_tts_mos}}
\vspace{-0.1cm}
\begin{tabular}{lcc|cc}
\toprule
\multirow{2}{*}{\textbf{Methods}} & \multicolumn{2}{c|}{\textbf{LibriSpeech}} & \multicolumn{2}{c}{\textbf{RAVDESS}} \\
\cmidrule(lr){2-3} \cmidrule(lr){4-5}
& \textbf{MOS-N$\uparrow$} & \textbf{MOS-S$\uparrow$} & \ {\textbf{MOS-N$\uparrow$}} & \ {\textbf{MOS-S$\uparrow$}} \\
\midrule
Groundtruth & 3.99 & 3.58 & 4.35 & 4.60\\
YourTTS$^{\lozenge}$ & 2.26 & 2.15 & 2.42 & 1.56\\
VALL-E* &  2.49 & 2.81 & 2.19 & 1.95\\
\midrule
Hierarchical LM & 3.19 & 3.31 & 3.24 & 3.19 \\
UniCodecLM & \textbf{3.36} & \textbf{3.59} & \textbf{3.74} & \textbf{3.72}\\  
\bottomrule
\end{tabular}
\end{center}
%\vspace{-0.2cm}
\end{table}

\subsection{Prosody-aware Speech
Language Modeling Evaluation\label{exp_D}}
\textbf{Zero-shot Text-to-Speech Synthesis} We use two datasets of different scales as our training data, LibriTTS\cite{Libritts}, which includes 585 hours of speech, and the medium split of the Libri-Light\cite{Librilight}, containing around 6K hours of speech. For the unlabeled Libri-Light data, we use a public wav2vec2-based ASR model\footnote{\url{https://huggingface.co/jonatasgrosman/wav2vec2-large-xlsr-53-english}} to get the transcripts,  
% Following VALL-E\cite{valle}, we conduct zero-shot TTS evaluation on the 4-10s subset from LibriSpeech test-clean set. 
% \textcolor{red}{We randomly select 3s clips as prompts from the same speaker's utterance.} 
\ {and each utterance is cropped to 3 to 20 seconds with a variable length. The TTS models are trained on 16 V100 GPUs with a batch size of 40 seconds per GPU.}
We compare our UniCodec token-based TTS system UniCodecLM with YourTTS\cite{yourtts}, VALL-E\cite{valle} and USLM\cite{SpeechTokenizer}. For evaluation, we employ two benchmark test sets: 1) LibriSpeech \textit{test-clean} subset. For a fair comparison, we evaluate on the 4-10\textit{s} subset only and randomly select one clip from the same speaker as the prompt following VALL-E.  2) RAVDESS, an emotional speech dataset. 
We use this test set to evaluate whether the TTS systems could preserve the desired emotion, prosody and timbre with high expressiveness. We evaluate the TTS systems from three aspects: naturalness (NISQA), intelligibility (WER) and speaker similarity (SPK). Each evaluation experiment runs five times and the average score along with the standard error is reported. 

As shown in Table \ref{Tab_tts}, with only 585 hours of parallel data, our UniCodecLM exhibits superior performance than existing methods trained with much larger training data, with higher intelligibility and naturalness, highlighting the effectiveness of our UniCodec token based approach in preserving both linguistic and paralinguistic information. Even lower WER is achieved when we increase the amount of data, showing the potential of our UniCodecLM. 
% It is noteworthy that our UniCodec utilizes significantly less data compared to the baselines (approximately tenfold less) , highlighting the effectiveness and precision of our token-based approach in capturing linguistic information. Additionally, our UniCodec also demonstrates superior naturalness and speaker similarity. 
Compared with Hierarchical LM using mHuBERT tokens as intermediate representations, our approach incorporates residual tokens $\bm{P}$ as target, as stated in Eq. \ref{Eq_tts}, 
%and global tokens $\bm{G}$ from the prompt, 
producing speech with more natural prosody with higher NISQA and better speaker attribute preservation. We also found in our experiments that the acoustic generation process of Hierarchical LM is not very stable and multiple inferences are often required to achieve satisfactory results, resulting in large standard errors of WER, as shown in Table \ref{Tab_tts}. In contrast, our UniCodecLM shows good robustness. This suggests that moving prosody generation to the first stage benefits the stability and robustness of the acoustic generation. Additionally, the joint optimization of semantic and residual token with the generative decoder in UniCodec token learning also facilitates stable acoustic generation. 
% \ {Additionally, higher speaker similarity is observed with higher SPK-O, indicating the effectiveness of UnicodecLM in modeling speaker-related information.}

Further exploration about the effectiveness of the disentanglement modeling of $\bm{S}$ and $\bm{P}$ in our UniCodec is conducted by comparing with the non-disentanglement variant of UniCodec, UniCodec-w/o disen, \ {the corresponding TTS model of which is denoted as UniCodecLM-w/o disen in Table \ref{Tab_tts}}. We could observe a significant loss on both WER and NISQA by UniCodecLM-w/o disen, indicating that the disentangled information structure is more conducive to modeling both linguistic and paralinguistic information. 
We also compare with a variant that assigns equal weights to $\bm{S}$ and $\bm{P}$ prediction losses without prioritizing $\bm{S}$ in TTS, denoted as UniCodecLM-w/o weighted in Table \ref{Tab_tts}. We observe slightly higher WER without weighting, indicating the effectiveness of prioritizing the semantic prediction.
% underscoring the effectiveness and potential of our decoupled modeling paradigm.
 
 Table\ref{Tab_tts2} shows the results of the expressive TTS on RAVDESS dataset,  \ {where we also evaluate the emotional accuracy on the synthesized speech with a wav2vec2-based emotion recognition model\footnote{\url{https://huggingface.co/ehcalabres/wav2vec2-lg-xlsr-en-speech-emotion-recognition}} as the classifier.} We can see that our UniCodecLM consistently outperforms YourTTS and Hierarchical LM by a remarkable margin in terms of naturalness, emotional accuracy and speaker similarity, indicating that our UniCodecLM is effective in preserving paralinguistic information. Besides, we can see that even without information disentanglement (UniCodecLM-w/o disen), there is still a significant improvement in naturalness and emotional accuracy compared to Hierarhical LM, which highlights the importance of incorporating prosodic information into speech language modeling.

\ {In addition to above objective metrics, we also employ mean opinion score (MOS) through human evaluations, as shown in Table \ref{Tab_tts_mos}, where 10 participants are invited to assess the naturalness of the synthesized speech (MOS-N) and its similarity to the prompt (MOS-S) on 6 samples from LibriSpeech \textit{test-clean} subset and 6 from the RAVDESS test set. It is shown that our UniCodecLM outperforms all other baselines in both naturalness and the similarity to prompts, which complies with observations from Table \ref{Tab_tts} and \ref{Tab_tts2}. We also observe that a much higher gain over Hierarchical LM in MOS-S is achieved on the RAVDESS dataset, indicating that for challenging cases with rich paralinguistic information such as emotional speech, relying solely on in-context learning to obtain such information from prompts is not as efficient as the explicit modeling in UniCodec. Our model demonstrates a higher capacity for capturing paralinguistic information, resulting in more expressive speech synthesis.} 
 
\begin{table}[!t]
\begin{center}
\caption{Evaluation of speech-to-speech translation quality. }
% \vspace{-0.1cm}
\begin{tabular}{lcccc}
\toprule
%\hline
\textbf{Methods} & \textbf{BLEU$\uparrow$} & \textbf{NISQA$\uparrow$}&\ {\textbf{MOS-N$\uparrow$}}&\ {\textbf{MOS-S$\uparrow$}} \\ %& SPK$\uparrow$ & Emotion$\uparrow$ \\
\midrule
%\hline
% TF-Codec &  - & - & - & - \\ 
mHuBERT-Vocoder  &  17.43 & 3.38& \ {4.23}&\ {1.29}\\ %& 0.05 & 0.27 \\
Hierarchical LM & \textbf{17.72} & 3.06 &\ {3.77}&\ {4.07}\\ % & 0.32 & 0.45 \\
UniCodecLM &  17.05 & \textbf{3.49} & \ {\textbf{4.37}}& \ {\textbf{4.11}}\\ %& 0.21 & 0.54 \\ 
%\hline
\bottomrule
\end{tabular}
\label{Tab_S2ST}
\end{center}
\end{table}

%\vspace{-0.2cm}
\textbf{Speech-to-Speech Translation} We train on the \textit{Hi-Res} subsets of the CVSS-T dataset\cite{cvss}, \ {which includes approximately 510K clips, each with a maximum duration of 20 and 17 seconds in source and target languages, respectively. We train translation models using 8 V100 GPUs with a batch size of 60.}
%\ {Instead of using the original target language speech, we 
%use an off-the-shelf TTS tool\footnote{https://github.com/yl4579/StyleTTS} to prepare synthetic target speech and %要不要说styletts生成的target呢
%apply the ASR model described in section \ref{exp_A} on the
%synthetic speech and filter out samples with word error rate (WER) greater than 80\%, which results in approximately 7.7k clips, each with a duration between 0 and 10 seconds. We train the translation models using 8 V100 GPUs with a batch size of 60.} 
We evaluate the synthesized speech in terms of naturalness (NISQA) and translation accuracy (ASR-BLEU) on the Spanish-English pair. \ {Subjective metrics MOS-N and MOS-S are also employed to evaluate the naturalness of the translated speech and the speaker similarity to the source language speech, where 10 participants are invited to assess 8 randomly selected samples.} Besides the Hierarchical LM, we also compare with another baseline mHuBERT-Vocoder, which uses mHuBERT as the target unit and a HiFi-GAN\cite{hifigan} unit-to-waveform vocoder to get the translated speech, similar to \cite{DS2ST}. This vocoder is language-specific and trained on single-speaker datasets. As shown in Table \ref{Tab_S2ST}, UniCodecLM achieves a comparable BLEU score to mHuBERT, showing the semantic representation capability of our tokens. \ {As mHuBERT token discards most of the paralinguistic information, mHuBERT-Vocoder fails to preserve the speaker characteristics in translated speech, resulting in a very low MOS-S. Comparatively, UniCodecLM explicitly models speaker characteristics in the global token $\bm{G}$, achieving the best MOS-S among all.}
Moreover, more natural prosody of the translated speech is observed in our UniCodecLM, with much higher NISQA \ {and MOS-N.}  
% which indicates the generated speech preserves paralinguistic information of the source well during translation. 
% We observed that the English speech generated by the Hierarchical LM exhibits a fast and unnatural rhythm, resulting in a much lower ASR-BLEU score and NISQA score. 
We observe in our experiment that the English speech generated by the Hierarchical LM often carries a Spanish accent due to the in-context learning from speech prompts in source language during acoustic generation. Spanish typically has a fast pace with many syllables and this pace is leaked from prompt into the target part, making it sound unnatural. Nevertheless, with explicit modeling of paralinguistic information in our UniCodecLM, the prosody is simultaneously translated with $\bm{S}$ during translation without accent issues,
% We believe that this is attributed to the domain gap between the prompt (source Spanish speech) and the target (English) during acoustic generation. Spanish typically has a fast pace with many syllables and this pace is leaked from prompt into the target part, making it sounding unnatural. 
which also indicates the necessity of our paralinguistic modeling in the S2ST task.

\section{Conclusion}
We present UniCodec, a universal speech token learning paradigm, that effectively models both linguistic and paralinguistic information in speech. The learned tokens are disentangled, compact and paralinguistic-rich, facilitating both understanding with paralinguistic clues and high-quality speech synthesis with prosodic and paralinguistic details well preserved. Extensive experiments demonstrate its effectiveness in various tasks. This work is a good exploration of what tokens can benefit speech language modeling and the proposed UniCodec token and the decoder serves as the foundation to build better paralinguistic-aware speech language models, especially for scenarios where paralinguistic information is essential such as spoken dialogue. This work is just a research project. As UniCodec itself can change the speaker attribute in voice conversion, it may have potential risks in misuse of the method such as impersonating a specific speaker. A protocol should be in place to ensure the speakers’ approval for using their speech or voice and altering the attributes. We will also put Microsoft AI Principles\footnote{\href{https://www.microsoft.com/ai/responsible-ai}{https://www.microsoft.com/ai/responsible-ai}} into practice in our future research.

\appendix
%\subsection{Global encoder structure}
\vspace{-0.2cm}
\begin{table}[!h]
\caption{Global encoder structure.}
%\label{sample-table}
%\vskip 0.15in
\vspace{-0.2cm}
\begin{center}
\begin{small}
\begin{tabular}{lp{4cm}}
\toprule
\textbf{Layers} & \textbf{Params} \\
\midrule
STFT & Hanning window, a window length of 40 ms, a hop length of 10 ms\\
Convolution & 5 layers, each kernel size 3 $\times$ 5, channels (16, 32, 32, 64, 64), strides ((1,1), (1,2), (1,4), (1,4), (1,2))\\
Reshape    &  T$\times$5$\times$64 to T$\times$320 \\
Temporal-filtering block \cite{Disen-TF-Codec} & 4 blocks, each kernel size 5 with increasing dilations and a channel dimension of 512 \\
Linear    &  input dimension 320, output dimension 512\\
Transformer encoder block    & 2 layers, a hidden dimension of 512, a feed-forward dimension of 2048, 8 attention heads \\
Multi-head attention block    &  a hidden dimension of 512, 8 attention heads\\
Linear    &  input dimension 512, output dimension 256 \\
\bottomrule
\end{tabular}
\end{small}
\end{center}
\vskip -0.1in
\end{table}

\bibliography{ref}
\bibliographystyle{IEEEtran}

\newpage

 

% \end{IEEEbiography}



\vfill

\end{document}